\documentclass{aa}
\def\folio{\ifnum\pageno=1\nopagenumbers\else\number\pageno\fi}

\def\lax    {\ifmmode{_<\atop^{\sim}}\else{${_<\atop^{\sim}}$}\fi}
\def\gax    {\ifmmode{_>\atop^{\sim}}\else{${_>\atop^{\sim}}$}\fi}
\newbox\grsign      \setbox\grsign=\hbox{$>$} 
\newdimen\grdimen   \grdimen=\ht\grsign
\newbox\simgreatbox \setbox\simgreatbox=\hbox{\raise.5ex\hbox{$>$}\llap
                        {\lower.5ex\hbox{$\sim$}}}\ht1=\grdimen\dp1=0pt
\newbox\simlessbox  \setbox\simlessbox =\hbox{\raise.5ex\hbox{$<$}\llap
                        {\lower.5ex\hbox{$\sim$}}}\ht2=\grdimen\dp2=0pt

\newbox\grsign \setbox\grsign=\hbox{$>$} \newdimen\grdimen \grdimen=\ht\grsign
\newbox\laxbox \newbox\gaxbox
\setbox\gaxbox=\hbox{\raise.5ex\hbox{$>$}\llap
     {\lower.5ex\hbox{$\sim$}}}\ht1=\grdimen\dp1=0pt
\setbox\laxbox=\hbox{\raise.5ex\hbox{$<$}\llap
     {\lower.5ex\hbox{$\sim$}}}\ht2=\grdimen\dp2=0pt
\def\gax{\mathrel{\copy\gaxbox}}
\def\lax{\mathrel{\copy\laxbox}}
\def\boxit#1    {\vbox{\hrule\hbox{\vrule\kern3pt
                  \vbox{\kern3pt#1\kern3pt}\kern3pt\vrule}\hrule}}
\def\h      {\ifmmode{^{\rm h}}\else{$^{\rm h}$}\fi}
\def\m      {\ifmmode{^{\rm m}}\else{$^{\rm m}$}\fi}
\def\s      {\ifmmode{^{\rm s}}\else{$^{\rm s}$}\fi}
\def\decas    {\ifmmode{{\rlap.}{''}}\else{${\rlap.}{''}$}\fi}
\def\mum     {\ifmmode{\mu{\rm m}}\else{$\mu{\rm m}$}\fi}
\def\s      {\ifmmode{^{\rm s}}\else{$^{\rm s}$}\fi}
\def\deg      {\ifmmode{^{\circ}}\else{$^{\circ}$}\fi}
\def\as     {\ifmmode {\rlap.}$\,$''$\,$\! \else ${\rlap.}$\,$''$\,$\!$\fi}
\def\decsec  {\ifmmode {\rlap.}$\,$^{s}$\,$\! \else ${\rlap.}$\,$^{s}$\,$\!$\fi}\def\decs  {\ifmmode {\rlap.}$\,$^{s}$\,$\! \else ${\rlap.}$\,$^{s}$\,$\!$\fi}

\def\kms    {\ifmmode{{\rm km~s}^{-1}}\else{km~s$^{-1}$}\fi}

\def\Lsun   {$L_{\odot}$}

\def\Msun   {$M_{\odot}$}
\def\Mspy   {\ifmmode {M_{\odot} {\rm yr}^{-1}} \else $M_{\odot}$~yr$^{-1}$\fi}
\def\Mdot   {\ifmmode {\dot M} \else $\dot M$\fi}
\def\mhd    {\ifmmode {n_{{\rm H}_2}} \else $n_{{\rm H}_2}$\fi}
\def\mhcd   {\ifmmode {N_{{\rm H}_2}} \else $N_{{\rm H}_2}$\fi}

\def\El      {\ifmmode{E_{\ell}}\else{$E_{\ell}$}\fi}
\def\beam    {\ifmmode{\theta_{\rm B}}\else{$\theta_{\rm B}$}\fi}
\def\mjyb   {\ifmmode {{\rm mJy~beam}^{-1}} \else{mJy~beam$^{-1}$}\fi}
\def\mujyb   {\ifmmode {\mu{\rm Jy~beam}^{-1}} \else{$\mu$Jy~beam$^{-1}$}\fi}
\def\Trot   {\ifmmode{T_{\rm rot}}\else$T_{\rm rot}$\fi}    
    
\def\Teff   {\ifmmode{T_{\rm eff}}\else$T_{\rm eff}$\fi}

\def\ITRS   {\ifmmode{\smallint {\rm T}_{R}^{*}dv}\else{$\smallint 
{\rm T}_{R}^{*}dv$}\fi}
\def\ITRS   {\ifmmode{\smallint {\rm T}_{R}^{*}dv}\else{$\smallint 
{\rm T}_{R}^{*}dv$}\fi}
\def\ITAS   {\ifmmode{\smallint {\rm T}_{A}^{*}dv}\else{$\smallint 
{\rm T}_{A}^{*}dv$}\fi}

\def\lefttitle#1  {\noindent \hangindent=18.0pt \hangafter=1 {#1} \par}
\def\vol#1  {{\bf {#1}{\rm,}\ }}
\font\tenssb=cmssbx10
\font\tenbf=cmbx10
\font\sevenbf=cmbx8
\font\fivebf=cmbx6
\def\unetdemi    {\smallskipamount=6pt plus2pt minus2pt
                  \medskipamount=12pt plus4pt minus4pt
                  \bigskipamount=24pt plus8pt minus8pt
                  \normalbaselineskip=16pt plus0pt minus0pt
                  \normallineskip=2pt
                  \normallineskiplimit=0pt
                  \jot=6pt
                  {\def\smallskip {\vskip\smallskipamount}}
                  {\def\medskip   {\vskip\medskipamount}}
                  {\def\bigskip   {\vskip\bigskipamount}}
                  {\setbox\strutbox=\hbox{\vrule 
                    height17.0pt depth7.0pt width 0pt}}
                  \parskip 12.0pt
                  \normalbaselines}
\def\smallerspace {\smallskipamount=3pt plus0pt minus0pt
                  \medskipamount=6pt plus0pt minus0pt
                  \bigskipamount=10.5pt plus0pt minus0pt
                  \normalbaselineskip=10.5pt plus0pt minus0pt
                  \normallineskip=1pt
                  \normallineskiplimit=0pt
                  \jot=3pt
                  {\def\smallskip {\vskip\smallskipamount}}
                  {\def\medskip   {\vskip\medskipamount}}
                  {\def\bigskip   {\vskip\bigskipamount}}
                  {\setbox\strutbox=\hbox{\vrule 
                    height8.5pt depth3.5pt width 0pt}}
                  \parskip 0pt
                  \normalbaselines}
\def\memospace    {\smallskipamount=4pt plus1pt minus1pt
                  \medskipamount=6pt plus2pt minus2pt
                  \bigskipamount=14pt plus6pt minus6pt
                  \normalbaselineskip=14pt plus0pt minus0pt
                  \normallineskip=1pt
                  \normallineskiplimit=0pt
                  \jot=4pt
                  {\def\smallskip {\vskip\smallskipamount}}
                  {\def\medskip   {\vskip\medskipamount}}
                  {\def\bigskip   {\vskip\bigskipamount}}
                  {\setbox\strutbox=\hbox{\vrule 
                    height17.0pt depth7.0pt width 0pt}}
                  \parskip 2.0pt
                  \normalbaselines}
\def\memowidespace    {\smallskipamount=5pt plus1pt minus1pt
                  \medskipamount=7.5pt plus2pt minus2pt
                  \bigskipamount=17.5pt plus6pt minus6pt
                  \normalbaselineskip=17.0pt plus0pt minus0pt
                  \normallineskip=1.25pt
                  \normallineskiplimit=0pt
                  \jot=5pt
                  {\def\smallskip {\vskip\smallskipamount}}
                  {\def\medskip   {\vskip\medskipamount}}
                  {\def\bigskip   {\vskip\bigskipamount}}
                  {\setbox\strutbox=\hbox{\vrule 
                    height21.25pt depth8.75pt width 0pt}}
                  \parskip 2.5pt
                  \normalbaselines}
\usepackage{graphicx}
\usepackage{natbib}
\bibpunct{(}{)}{;}{a}{}{,} 
\usepackage{ulem}
      \def\new#1 {{\bf #1 }}
      \def\cut#1 {\sout{#1} }

\usepackage{longtable}
\usepackage{amsmath}

      \def\new#1 {{\bf #1 }}
      \def\cut#1 {\sout{#1} }

\def\beq{\begin{equation}}
\def\eneq{\end{equation}}
\def\simgt{\lower.5ex\hbox{$\; \buildrel > \over \sim \;$}}
\def\simlt{\lower.5ex\hbox{$\; \buildrel < \over \sim \;$}}

\begin{document}

\title{SiO maser emission from red supergiants across the Galaxy: I. Targets in massive star clusters}

\author{L. Verheyen\thanks{Member of the International Max Planck Research School (IMPRS) for Astronomy and
Astrophysics at the Universities of Bonn and Cologne}
\and
M. Messineo
\and
K. M. Menten
}
\offprints{K. M. Menten}

\institute{Max-Planck-Institut f\"ur Radioastronomie,
Auf dem H\"ugel 69, D-53121 Bonn, Germany\\
\email{verheyen, messineo, kmenten@mpifr-bonn.mpg.de}
}

\date{Received    2011/ Accepted }
\titlerunning{Red Supergiant SiO Masers in the Galaxy}

\authorrunning{Verheyen et al.}

\abstract
{}
   {Red supergiants (RSGs) are among the most luminous of all stars, easily detectable in external galaxies, and
   may ideally serve as kinematic tracers of Galactic structure. Some RSGs are surrounded by circumstellar
   envelopes detectable by their dust and molecular and, in particular, maser emission. This study consists of a  
   search for maser emission from silicon monoxide (SiO) toward a significant number of RSGs that are members 
   of massive stellar clusters, many of which have only recently been discovered. 
   Further, we aim to relate the occurrence of maser action to properties of the host stars. 
}
   {Using the IRAM 30 meter telescope, we searched for maser emission in the $J = 2 -1$ rotational transition within the first vibrationally excited state of SiO toward a sample of 88 RSGs.
}
   {With an average rms noise level of 0.25 Jy, we detected maser emission in 15\% of the sample, toward most of the sources for the first time in this transition. The peak of the emission provides accurate radial velocities for the RSGs. The dependence of the detection rate on infrared colors  supports a radiative pumping mechanism for the SiO masers.
}
{}


\maketitle

\section{\label{introduction}Introduction}

Near the end of their lifetimes, stars of moderately high mass (10 $M_\odot$ $< M_* <$ 30 $M_\odot$) go through the red supergiant 
evolutionary phase. Red supergiants (RSGs) are burning helium in their core, have luminosities more than $10^5$ times that of the Sun and,  
since their effective temperature is low ($\sim3000$--4000 K),  are intrinsically luminous in the near-infrared range (NIR) \citep{levesque10}.  
Studying RSGs and, in particular, their mass-loss history, is important for several reasons. Given their luminosities, RSGs may be  
recognized as individual objects in external galaxies and serve as tracers of young populations and as distance indicators. 
They contribute to the enrichment of the interstellar medium by losing mass at a high rate (up to several times $10^{-4}$ $M_\odot$ yr$^{-1}$), and by eventually exploding as core-collapse supernovae. RSG stars are often surrounded by circumstellar envelopes (CSEs).  
Maser emission may arise from these envelopes, revealing stellar line-of-sight velocities to within a few \kms\ \citep{jewell91}. 
Frequently detected maser lines are from OH at 1.612 GHz, H$_2$O at 22.2 GHz, and SiO at 43 GHz and 86 GHz. At higher frequencies, also a number of (sub)millimeter transitions have been detected \citep[see, e.g. ][]{Humphreys2007, Menten2008}.

Some fundamental issues limit the use of RSGs as well-calibrated tracers of young stellar populations. We know only about 500 RSGs \citep[e.g.][]{messineo11,figer06,davies07}, and their properties (stellar and circumstellar) are still poorly understood. Interstellar extinction and poor knowledge on stellar distances render the identification of  RSGs difficult. Important open questions are those of the relations (i) between the maximum luminosity of a given star and its main-sequence mass and (ii) between mass-loss rate, $\dot{M}$, 
and stellar luminosity, $L_*$ and  initial metallicity, $Z$. Possible stellar rotation complicates these relations 
\citep{meynet00}. Periodic stellar light variations and  circumstellar extinction make it difficult to distinguish between  
massive RSGs and intermediate age mass-losing late-type stars  \citep{comeron04, messineo05, messineo11}.

All these uncertainties limit our understanding of these stars' CSEs. In particular, it is unclear 
why some stars exhibit maser emission in a certain species and others don't. Maser occurrence is certainly not only a function 
of envelope temperature and mass-loss rate alone; other parameters, like elemental abundances and stellar luminosity may play an important role \citep{habing96}.

Large-scale searches for maser emission toward evolved late-type stars (mostly Mira-type long period variables and OH/IR stars) have been 
carried out, successfully yielding radial 
velocity measurements, and new views on the structure of the Milky Way, in particular on its bar and bulge  \citep{sevenster99,Deguchi2002,habing06}. 
The power of velocity determinations via  observations of SiO maser lines  was demonstrated by \citet{messineo02,messineo04,messineo05} and \citet{habing06}, 
who used the IRAM 30m telescope to find  86 GHz SiO maser emission toward 271 color-selected
AGB stars in the Inner Galaxy with a detection rate higher than 60\%. With ages of up to a few tens of Myr, the much younger RSGs, on the other hand, are found in the general vicinity of their birth places.

Recently, trigonometric parallaxes of maser sources in star-forming regions and toward a few RSGs located in the  spiral arms of the Milky Way have been measured with the NRAO
Very Long Baseline Array (VLBA) and VLBI Exploration of Radio Astrometry (VERA) with a precision of 10\%. These accurate distances  are  leading to a revised  view of Galactic structure \citep{reid09,choi08, asaki10}. 
RSGs are found in the spiral arms and their associated masers are thus excellent potential targets for such studies.
In fact, one purpose of the present work is to increase the number of RSG/SiO maser sources suitable for Very Long Baseline Interferometry (VLBI).

In order to use RSGs as kinematic probes, it is necessary to significantly increase the number of such objects with known associated maser emission. Strong masers (with hundreds to $>$ thousand Jy intensities) were first detected in nearby ``classical'' OH/IR RSGs (VY CMa, VX Sgr, NML Cyg, and also S Per) in the early 1970s \citep{buhl74,kaifu75}. Searches for RSG masers in the far reaches of the Galaxy require sensitive searches with few- to even sub-Jy sensitivity.

In this paper, we present  a search for 86 GHz SiO maser emission toward 88 known RSG stars associated with open clusters.

In Sect. \ref{rsgs}, we report  recent discoveries of massive clusters that have RSGs amongst their members (the so named Red Supergiant Clusters, RSGCs), and in Sect. 
\ref{maserbasics} give a brief summary of SiO maser emission and the motivation for its study. 
The observations are described in Sect. \ref{observations}. General results are presented in Sect. \ref{results}, discussed, 
cluster by cluster, in Sect. \ref{individual} and analyzed in Sect. \ref{analysis}. Some conclusions are presented in Sect. \ref{conclusions}.

\section{Red supergiants in stellar clusters\label{rsgs}}

Ideally, one wishes to observe RSG stars in clusters, because membership in a stellar cluster guarantees 
similar distance, age and metallicity  for all its constituents, including the RSGs. The RSGs' distinct properties, compared to other cluster members, will thus
only depend on their large initial masses. Due to the steepness of the initial mass function
stellar clusters rich in RSGs must naturally be very massive, and
have an age from about 5 to 30 Myr \citep{meynet00}.

Only 15 massive clusters (mass $> 10^4$ \Msun) are currently known in the Milky Way
\citep[the most recently found ones reported by ][]{messineo09,negueruela10,negueruela11,messineo11}. They are all located at the near-side of the Galaxy,  
suggesting that their census is highly incomplete  \citep{messineo09}. A few clusters, e.g. the Westerlund I, Arches, and Quintuplet clusters \citep{westerlund61,cotera96,figer99,figer02,figer04,clark05} were known to be massive since the 1990s or even earlier and have been studied at various wavelengths. However, the Arches and the Quintuplet may not be representative of massive clusters elsewhere in the Milky Way  because of the peculiar physical conditions in the Galactic center region. Thanks to the new generation of near- and mid-infrared surveys \citep[e.g. DENIS, 2MASS, GLIMPSE, ][]{denis2,cutri03,benjamin03}, recently several more massive Galactic stellar clusters containing RSGs have been discovered around $l \sim 25^\circ$, at the base of the Scutum-Crux arm. In the Red SuperGiant Cluster 1 (RSGC1) a total of 14 RSGs have been spectroscopically confirmed \citep{figer06,davies08}. Only a few hundred parsecs from the RSGC1, another massive cluster was found. RSGC2, also called the Stephenson 2 cluster \citep{stephenson90}, contains 26 RSGs \citep{davies07}. A nearby third massive cluster (RSGC3) with 16 RSGs was discovered by \citet{clark09} and \citet{alexander09}. Recently, two other clusters rich of RSGs were identified  in the same direction \citep{negueruela10,negueruela11}. The Alicante 8 or RSGC4 contains 13 candidate RSG members, while the RSGC5 cluster contains seven RSGs. Since the latter two  clusters were not yet discovered at the time of the observations presented in this paper, they were not considered in our search. The total number of confirmed RSG stars in these five clusters amounts to 64, increasing the number of known RSGs by at least 15\%. Before this, only the NGC 7419 cluster was known to contain five RSGs \citep{beauchamp94,caron03}. All 
five RSGCs are concentrated within a few degrees of longitude (25$^\circ < l < 29^\circ$), likely at similar distance (5--6 kpc). The
location of this remarkable burst of star formation coincides with the near endpoint of the Galactic bar \citep{davies09}, where also the W43 ``mini starburst'' region is found \citep{Nguyen2011}. These clusters offer an extraordinarily important natural laboratory in which to study the evolution of RSGs and the properties of their envelopes. Altogether, we now have a sample of more than 119 Galactic RSGs in clusters at our disposal \citep{messineo11}. Here
we present a search for 86 GHz SiO maser emission toward RSG stars. 

The RSGCs are at distances of approximately
3.5 kpc from the Galactic center, all are at kinematically interesting locations, namely at the corotation radius of the Galactic bar \citep{habing06},
at the base of the Scutum-Crux arm, or near an endpoint of the Galactic bar.
Studies of their  properties and  kinematics, e.g. via SiO maser emission, will place constraints on
Galactic structure and evolution models and address, e.g., questions how the Galactic bar and the spiral arms are dynamically related, which is still unclear \citep{bissantz03}. 

\section{SiO masers in red supergiants\label{maserbasics}}

Analysis of existing photometric data (from the 2MASS, MSX, Spitzer/GLIMPSE 
databases) has shown that many RSGs show infrared excesses, caused by their CSEs, and therefore high mass-loss rates \citep[e.g.][]{figer06,davies07}.  Typical mass-loss rates range from 10$^{-8}$ to  10$^{-4}$ \Mdot\ per 
year \citep{verhoelst09, harwit01}. This makes them likely to exhibit maser emission, as, e.g., the SiO maser flux is known to correlate with the mid-infrared continuum flux density see Sect. \ref{maserpumping}.
Most commonly, RSGs show maser emission from OH (at 1.612 GHz), H$_2$O (22 GHz) and SiO 
(43 GHz and 86 GHz) \citep[see e.g.][]{habing96}. In this paper, we will focus on the 
86 GHz SiO $(v= 1, J=2-1)$ maser line.

\subsection{SiO maser pumping\label{maserpumping}}

The pumping mechanism for SiO masers is still under debate. Both radiative and collisional
pumping models have been proposed. The basic mechanism for SiO maser pumping is described by
\citet{kwan74}. They find that SiO population inversions can occur when vibrational transitions
become optically thick. Namely, in this case the vibrational de-excitation rates will be modified in
such way that they decrease with increasing rotational level $J$. The logical result is an
overpopulation of higher $J$ rotational levels, thus an inversion, as long as the pumping happens by
collisions or an indirect radiative route.

Models favoring either collisional or radiative pumping have been proposed by a number of authors. A
mechanism for pumping through collisions with H$_2$ molecules was first presented by
\citet{elitzur80}. Further modeling by, e.g., \citet{lockett92} and \citet{humphreys02} also
supported collisional pumping as the primary pumping mechanism for SiO masers. On the other hand,
contrasting studies find radiative pumping to be dominant
\citep[e.g.][]{bujarrabal87,bujarrabal94a,bujarrabal94b}. From the observational point of view, a
radiative pumping model explains the observed correlation between SiO maser intensity and the
infrared emission \citep[e.g.][]{bujarrabal87}, while the dependence on stellar pulsations could be
explained by collisional pumping \citep[e.g.][]{heske89}. The above models find that radiative pumping requires that the infrared (IR) flux at $8 \mu$m (the equivalent wavelength of the first excited rotational level of SiO) is greater than the flux of the maser line.

\subsection{Importance of SiO maser studies: obtaining accurate velocities}

To summarize the above: Finding new SiO maser-hosting RSGs associated with stellar clusters,  i.e., of stars in environments of similar age, metallicity, distance and interstellar extinction is useful for analyzing the occurrence of  SiO maser emission in RSGs, as a function of their luminosities and mass-loss rates \citep{davies08}. Very importantly, maser lines provide accurate radial velocities; the midpoint of an SiO maser spectrum  gives the stellar velocity with an accuracy of a few \kms\ \citep{jewell91}. In contrast, radial velocities from infrared  CO absorption at 2.29 $\mu$m in Mira stars may vary with the stellar pulsation up to 15 \kms\ \citep{scholz00}. We note, however, that for the RSGs for which we have both maser and CO absorption LSR velocity measurements, both values are quite similar; see Table \ref{rsgc1_vlsr}.

With accurate radial velocities of cluster members a determination of   dynamical cluster masses is possible. Furthermore, they allow a determination of  kinematic distances. However, such distances are based on models of Galactic rotation, i.e., only meaningful outside of the corotation radius and may be wrong due to peculiar motions \citep[see, e.g., ][]{Xu2006} . Given all this, a main motivation for this work is to find new samples of RSGs with SiO maser emission whose distances can be studied with VLBI. The distances thus obtainable, typically with uncertainties of 10\% at 10 kpc, will greatly improve our knowledge of massive cluster properties and place more stringent constraints on Galactic structure.


\section{Observations\label{observations}}

\subsection{Pico Veleta observations}

We observed the SiO $v$=1, $J$=2--1 maser transition at 86.24335~GHz in 88 sources with the
IRAM 30-m telescope  on Pico Veleta, Spain, during 3 epochs in 2006 September/October and 2008 May
and August (Table \ref{dates}). The A100 and B100 receivers were used, connected to the
256-channel low (1 MHz) resolution filterbank with 3.5~km~s$^{-1}$  spectral resolution, as well
as the VESPA autocorrelator with 240 MHz bandwidth  (1600 \kms) and a resolution of 40 kHz, corresponding 
to an adequate 0.14 km~s$^{-1}$.
We used a total power observing mode with integration times between 5 and 15 minutes per source,
employing wobbler  switching with a wobbler  throw of $100''$--$120''$. This gives an rms noise around 20--65~mK in 
a corrected antenna temperature, $T^*_A$,
scale, equivalent to 0.12--0.54~Jy, assuming a conversion factor from antenna temperature to
flux density of 6 Jy/K. The system temperatures ranged from 90  to 300~K. We conducted telescope pointing checks every 1.5--3~h and, based on these, estimate that  pointing errors are smaller than 3\arcsec, much smaller
than the 29\arcsec\ FWHM beam width. .

\begin{table}[tb]
\begin{center}
\caption{\label{dates}IRAM 30m observing dates}
\begin{tabular}{ll}
 \hline \hline
Epoch    & Dates  \\
\hline
1   & 2006 September 29-- October 2  \\
2   & 2008 May 21--23\\
3   & 2008 August 9--11  \\
\noalign{\smallskip}
 \hline
 \noalign{\smallskip}
 \end{tabular}
\end{center}

\end{table}

\subsection{Data reduction}

We reduced the data with the CLASS program, which is part of the GILDAS software package\footnote{http://www.iram.fr/IRAMFR/GILDAS}. Special attention was paid to the possible
confusion with the interstellar H$^{13}$CN $J = 1 - 0$ line, which has 3 hyperfine transitions in the 240 MHz VESPA
band \citep{messineo02}. We detected the H$^{13}$CN line in all of the Galactic  center sources
from the  \citet{blum03} sample.

\begin{table*}[t]
\begin{center}
\caption{\label{sio_table_detections}SiO~$\textit{v} = 1; J = 2-1$ detections toward red supergiants.}
\begin{tabular}{lccrcrcrc}
 \hline \hline
Object  & $\alpha_{J2000}$  & $\delta_{J2000}$  & $S$   & rms  &$\textit{v}_{\rm LSR}$ & $\textit{v}$-range  &  $\int S~d$\textit{v} & Ref. \\
  &   &  & [Jy]  & [Jy]  & [km s$^{-1}$] & [km s$^{-1}$]  & [Jy km s$^{-1}$] \\
\hline
RSGC1\\
\hline
FMR2006 1   & 18:37:56.29  & $-$06:52:32.2  & 1.2  & 0.20  & 116  & [112,129]  & 8.1$\pm$3.0   & (1) \\ 
FMR2006 3   & 18:37:59.73  & $-$06:53:49.4  & 0.6  & 0.18  & 112  & [110,117]  & 2.4$\pm$1.2   & (1) \\ 
FMR2006 4   & 18:37:50.90  & $-$06:53:38.2  & 1.1  & 0.18  & 121  & [115,130]  & 6.1$\pm$1.8   & (1) \\ 
FMR2006 13  & 18:37:58.90  & $-$06:52:32.1  & 1.6  & 0.26  & 113  & [108,128]  &14.7$\pm$3.0   & (1) \\
X 18     & 18:38:01.62  & $-$06:55:23.5  & 1.7  & 0.22  & 71   & [69,75]    & 3.1$\pm$1.2   & (1) \\
\hline
RSGC2\\
\hline
Stephenson 2 DFK 2   & 18:39:19.60  & $-$06:00:40.8  & 1.0  & 0.12  & 101  & [94,113]   &  6.9$\pm$1.8   & (2) \\ 
Stephenson 2 DFK 49  & 18:39:05.60  & $-$06:04:26.6  & 6.0  & 0.14  & 101  & [88,109]   & 59.0$\pm$1.8  & (2) \\ 
Stephenson 2 DFK 1   & 18:39:02.40  & $-$06:05:10.6  & 3.0  & 0.19  & 89   & [71,101]   & 30.5$\pm$3.0  & (2) \\
\hline
W Per        & 02:50:37.90  & +56:59:00.0    & 3.0  & 0.14  & $-$40& [-57,-29]  & 33.7$\pm$2.1  & (3) \\
S Per        & 02:22:51.70  & +58:35:12.0    & 91.1 & 0.21  & $-$41& [-50,-28]  & 716.4$\pm$2.7 & (3) \\
NGC 7419-MY Cep & 22:54:31.66  & +60:49:40.3 & 7.9  & 0.11  & $-$52& [-71,-42]  & 64.4$\pm$1.8  & (4) \\ 
MDI2009 GLIMPSE13 1  & 18:53:52.49  & +00:39:31.3    & 7.0  & 0.15  & 71   & [57,84]    & 67.9$\pm$2.1  & (5) \\
Cl Trumpler 27 1  & 17:36:10.12  & $-$33:29:40.6  & 50.3 & 0.27  & $-$17& [-25,9]    & 260.5$\pm$4.2 & (6) \\
IK Tau (Mira)& 03:53:28.84  & +11:24:22.6    & 496.2 & 0.27 & 33.3 & [30,36]    & 1074.0$\pm$1.5& (7) \\
U Her  (Mira)& 16:25:47.69  & +18:53:33.1    & 90.8 & 0.22  & $-$16& [-22,-12]  & 248.0$\pm$1.5  & (8) \\
\noalign{\smallskip}
\hline
\noalign{\smallskip}
\end{tabular}
\end{center}
\small
\noindent
\textbf{Notes.} Columns are (from left to right) designations of the targeted RSGs (names are taken from SIMBAD), right ascension, declination, peak flux density, 
rms noise, LSR velocity of peak flux density, LSR velocity range covered by maser emission (from a Gaussian fit), velocity-integrated 
flux density, and a reference number. \\
 \\
\textbf{References.} (1) RSGC1     \citep{figer06}; (2) RSGC2     \citep{davies07}; (3) Per OB1   \citep{pierce00}; 
(4) NGC 7419  \citep{caron03}; (5) GLIMPSE13 \citep{messineo09}; (6) Trumpler 27 \citep{massey01}; (7) IK Tau; (8) U Her.
\end{table*}

\addtocounter{table}{1}

\subsection{Target stars}

The targets of our 86 GHz observations are late-type stars that are located mostly in the directions of massive
clusters containing confirmed or candidate RSGs. Our sample is presented, with the outcome of our observations  in Tables \ref{sio_table_detections} and \ref{sio_table_nondetections}.
Table \ref{sio_table_detections} gives the coordinates of the targets toward which an
SiO maser was detected and the  line parameters (peak flux density, rms noise,  peak LSR velocity, velocity range of maser emission,
and velocity-integrated flux). Table \ref{sio_table_nondetections} reports the coordinates of the targets not
detected as well as the rms noise ($1\sigma$) of the spectra. Several of our target RSGs have already been found to show SiO maser action in
the $v = 1$ and/or 2, $J =1-0$ lines at 43 GHz  \citep[e.g.][]{nakashima06}, as well as  the  86 GHz $v = 1, J =2-1$ 
line \citep{haikala94} or the 43 GHz lines \textit{and} the 86 GHz line\footnote{While, in oxygen-rich 
evolved stars, the  SiO $v = 1$ and $2, J =1-0$ lines have comparable intensity, the $v = 2, J =2-1$ is always much weaker than the $v = 2, J =1-0$ line
and has not yet been reported in a RSG  \citep{bujarrabal96}.} 
 \citep{deguchi10}. Since SiO maser emission 
is often variable \citep[e.g][]{alcolea90,alcolea99}, a
second epoch of observations may reveal new sources as maser emitters even if previous searches had shown negative results.

\begin{figure*}[t!]
\begin{center}
\includegraphics[width=13.5cm,angle=0]{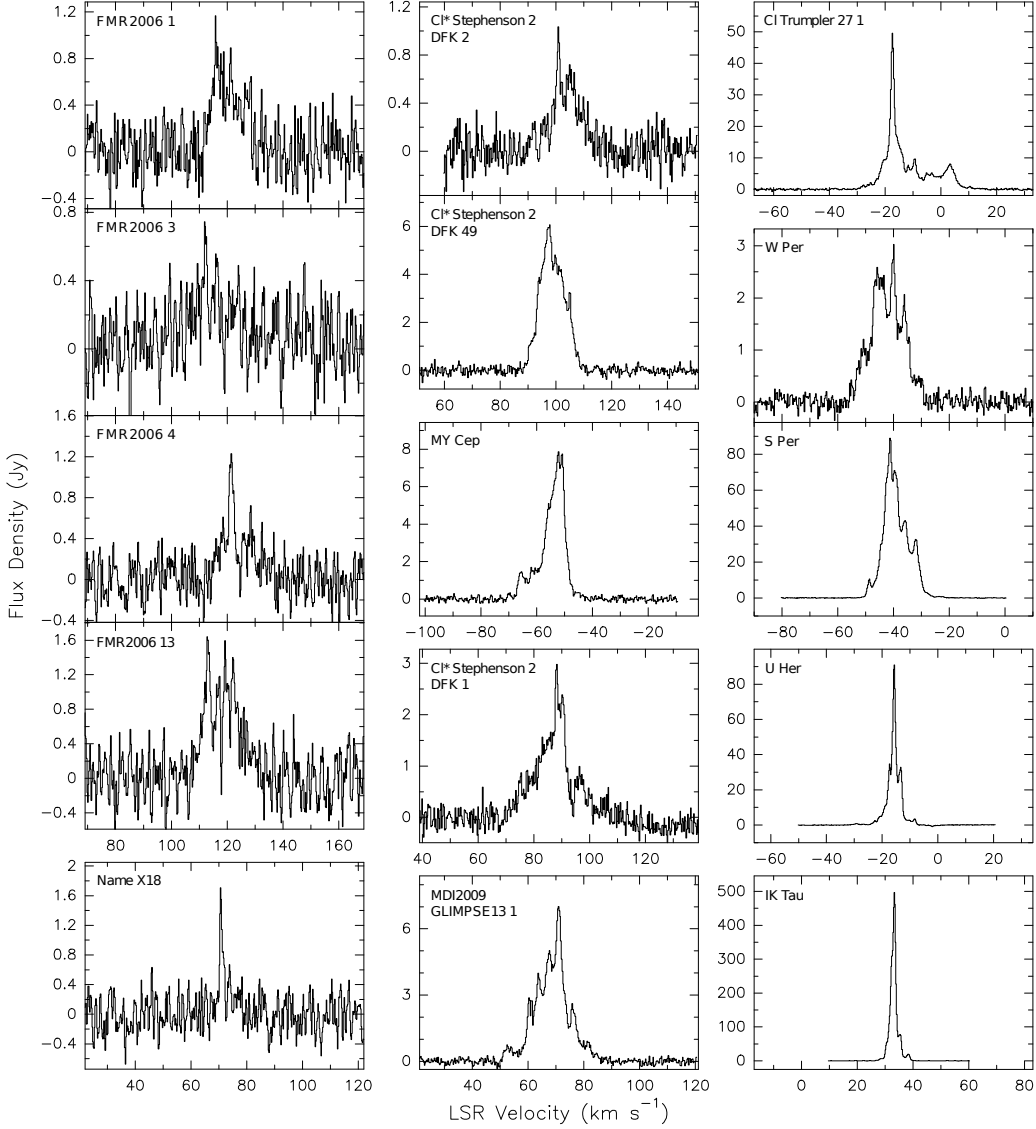}
\caption{\label{sio_detections_spectra}Spectra of the SiO $\textit{v} = 1$, $J =2-1$ transition toward the stars were the line was detected. All stars but X 18, U Her, and IK Tau are RSGs. The latter two are Mira variables and we show their much 
narrower spectra for comparison.}
\end{center}
\label{figdetections}
\end{figure*}

\section{General Results\label{results}}

We searched 88 (candidate) RSGs for 86 GHz SiO maser emission. We have detected SiO maser emission from RSGs in the RSGC1 and RSGC2, NGC 7419, GLIMPSE13, and Trumpler 27 clusters, and from two RSGs in the Perseus OB1 association, S Per and W Per, which were already known to harbor SiO masers \citep{kaifu75,jiang96}. In total, 13 out of the 88 observed RSGs were found to exhibit SiO maser emission in the $v=1, J=2-1$ transition at 86 GHz.

All spectra are presented in Figure \ref{sio_detections_spectra} and the observed line parameter are listed in Table \ref{sio_table_detections}. The
columns specify the RSG's name (taken from SIMBAD\footnote{http://simbad.u-strasbg.fr/simbad/}), its position by right ascension ($\alpha_{\rm J 2000}$) and declination ($\delta_{\rm J 2000}$), the peak flux density ($S$, in Jy),
the rms noise (rms, in Jy), the LSR velocity of the peak flux density ($\textit{v}_{\rm LSR}$, in \kms), the LSR velocity range covered by the maser emission
($v$-range, in \kms), the velocity-integrated flux density ($\int S\, dv$, in Jy \kms), and a reference number.  Table \ref{sio_table_nondetections}
lists upper limits for sources toward which no maser emission was found. The columns specify the RSG's name (taken from SIMBAD), its position by right
ascension ($\alpha_{\rm J 2000}$) and declination ($\delta_{\rm J 2000}$), and the rms noise (rms, in Jy).

Toward several of the  13 sources  in which we detect SiO  maser emission  SiO, OH and/or H$_2$ masers had been found previously. For these, Table \ref{tabliterature} gives an overview of all the maser detections.

\begin{table*}[t]
\begin{center}
\caption{\label{tabliterature} Previous maser observations}
\begin{tabular}{lcccccc}
\hline \hline
Name                    & H$_2$O & SiO & OH &$\textit{v}_{LSR}$ & References  & Comment \\
\hline
FMR2006 1               &        & SiO &    & +115.0    & (1)         &         \\ 
FMR2006 3               &        &     &    &           &             & New     \\     
FMR2006 4               &        & SiO &    & +121.5    & (1)         &         \\ 
FMR2006 13              &        & SiO &    & +116.5    & (1)         &         \\ 
X 18                &        & SiO &    & +74.3     & (1)         &         \\
Stephenson 2 DFK 2  & H$_2$O & SiO &    &+102.7     & (2)         &         \\
Stephenson 2 DFK 49 & H$_2$O & SiO &    &+98.6      & (2)         &         \\
Stephenson 2 DFK 1  & H$_2$O & SiO &    &+92.7      & (2)         &         \\
NGC 7419-MY Cep         & H$_2$O & SiO & OH & $-$55.8   & (3),(4),(5) &         \\
W Per                   & H$_2$O & SiO &    &-40.0      & (6)         &         \\
S Per                   & H$_2$O & SiO & OH &-37.0      & (4),(6)     &         \\
MDI2009 GLIMPSE13 1     &        & SiO &    & +72.3     & (2)         &         \\  
Cl Trumpler 27 1        &        & SiO &    &           & (7)         &         \\
IK Tau                  & H$_2$O & SiO & OH &   +35.0   & (4)         &         \\
U Her                   & H$_2$O & SiO & OH & $-$15.0   & (4)         &         \\
\hline \\
\end{tabular}
\end{center}
\small
\textbf{References:} (1) \citet{nakashima06}; (2) \citet{deguchi10}; (3) \citet{nakashima07}; (4) \citet{besanson90};
(5) \citet{takaba01}; (6) \citet{valdettaro01}; (7) \citet{hall90}.
\end{table*}

\begin{table*}[t]
\begin{center}
\caption{Comparison between RSG LSR velocities determined from SiO maser emission and IR CO absorption}\label{rsgc1_vlsr}
\begin{tabular}{lrrr}
  \hline
  \hline
  \noalign{\smallskip}
  Object & \multicolumn{3}{c}{\textit{v}$_{\rm LSR}$ [\kms]}\\
  \noalign{\smallskip}
         & \multicolumn{1}{c}{previous} & \multicolumn{1}{c}{this study} &  \multicolumn{1}{c}{CO} \\
         & \multicolumn{1}{c}{(43 GHz SiO)} & \multicolumn{1}{c}{(86 GHz SiO)} & \multicolumn{1}{c}{(2.29 $\mu$m)} \\
  \noalign{\smallskip}
  \hline
  \noalign{\smallskip}
  FMR2006 1               & 115\tablefootmark{a}$\;$ $\;$ $\;$    & 116$\;$ $\;$ $\;$ & 130\tablefootmark{g} \\
  FMR2006 2               & 120\tablefootmark{a}$\;$ $\;$ $\;$    &                   & 114\tablefootmark{g} \\
  FMR2006 3               &                                       & 112$\;$ $\;$ $\;$ & 127\tablefootmark{g} \\
  FMR2006 4               & 121\tablefootmark{a}$\;$ $\;$ $\;$    & 121$\;$ $\;$ $\;$ & 121\tablefootmark{g} \\
  FMR2006 13              & 116\tablefootmark{a}$\;$ $\;$ $\;$    & 113$\;$ $\;$ $\;$ & 125\tablefootmark{g} \\
  X 18                & 74\tablefootmark{a}$\;$ $\;$ $\;$     & 71$\;$ $\;$ $\;$  & \\
  \hline
  Stephenson 2 DFK 2  & 103\tablefootmark{b}$\;$ $\;$ $\;$    & 101$\;$ $\;$ $\;$ & 111\tablefootmark{h} \\
  Stephenson 2 DFK 49 & 101\tablefootmark{b}$\;$ $\;$ $\;$    & 101$\;$ $\;$ $\;$ & 109\tablefootmark{h} \\
  Stephenson 2 DFK 1  & 93\tablefootmark{b}$\;$ $\;$ $\;$     & 89$\;$ $\;$ $\;$  & \\
  \hline
  MY Cep                  & -54\tablefootmark{c}$\;$ $\;$ $\;$    & -52$\;$ $\;$ $\;$ & \\
  \hline
  W Per                   & -41\tablefootmark{d}$\;$ $\;$ $\;$    & -40$\;$ $\;$ $\;$ & \\
  S Per                   & -41\tablefootmark{e}$\;$ $\;$ $\;$    & -41$\;$ $\;$ $\;$ & \\
  \hline
  MDI2009 GLIMPSE13 1     & 69/72\tablefootmark{b}$\;$ $\;$ $\;$  & 71$\;$ $\;$ $\;$   & \\
  \hline
  Cl Trumpler 27 1        & -11\tablefootmark{f}$\;$ $\;$ $\;$    & -17$\;$ $\;$ $\;$ & \\
  \noalign{\smallskip}
  \hline
  \noalign{\smallskip}
\end{tabular}
\end{center}
\small
\textbf{Notes.} Columns are (left to right): RSG designation and LSR velocities  determined from the 
SiO $v=1, J = 1 - 0$ and $2 - 1$ lines at 43 and 86 GHz, respectively, and the 2.29 $\mu$m CO bandhead absorption. References are:\\
 \\
 \textbf{References.} \tablefoottext{a}{\citet{nakashima06}}; \tablefoottext{b}{\citet{deguchi10}}; \tablefoottext{c}{\citet{nakashima07}};
\tablefoottext{d}{\citet{jiang96}};  \tablefoottext{d}{\citet{kim10}}; \tablefoottext{f}{\citet{haikala94} -- velocity based on 86 GHz SiO maser line};  \tablefoottext{g}{\citet{davies08}}; \tablefoottext{h}{\citet{davies07}.}

\end{table*}

\section{Results on individual clusters\label{individual}}

In the following we discuss our targets and our observations cluster by cluster.

\subsection{RSGC1}

\citet{figer06} reported on a cluster containing 14 RSGs, the RSGC1 located at $l = 25\overset{\circ}{.}3, b
= -0\overset{\circ}{.}2$. High-resolution near-infrared spectroscopy confirmed their memberships
\citep{davies08}. The 14 RSG stars have spectral types between K2 and M6. The mean radial velocity of the
stars is  $123.0\pm1.0$ \kms, while the velocity dispersion is 3.7 \kms, which yields a dynamical mass of a
few times $10^4$ \Msun. With the derived kinematic distance of $6.6\pm0.9$ kpc, \citet{davies08} estimated
luminosities $\log(L/\textrm{\Lsun})$ from 4.87 to 5.45, which are consistent with a cluster age of  $12\pm2$
Myr and initial masses of 18 \Msun. We observed 8 of the 14 RSGs, for which we use the identification numbers introduced by \citet{davies08}. We also included the nearby source X 18. This bright  mid-infrared source was suggested to be a cluster member and possibly associated with the X-ray emitter AX J1838.0-0655 by \citet{figer06}. Toward this
source, located $2'$ south of the cluster,  SiO emission was found by \citet[][(see below)]{nakashima06}. A radial velocity of 74 \kms, significantly lower than that of the stellar cluster, however, suggests non-membership \citep{nakashima06}.
\citet{nakashima06} already searched for the SiO $v = 1$ and $v = 2$, $J =1-0$ transitions at 43
GHz in the RSGs of RSGC1. They detected maser emission from four of the 14 RSGs: FMR2006 1, FMR2006 2, FMR2006 4 and FMR2006 13. Only the latter was
detected in both transitions, the other three objects only showed the $v = 1$ line. Additionally, in a follow-up observation the 86 GHz $v=1, J=2-1$ SiO maser line
was detected in FMR2006 1 and FMR2006 2. On top of that, \citet{nakashima06} also found emission from both the $v = 1$ and $v = 2$ transitions 
toward X 18.

We detected 5 SiO masers in the direction of the RSGC1. We detected the $\textit{v} = 1$, $J =2-1$ transitions toward those objects found masing in the $J=1-0$ lines by \citet{nakashima06}, except for FMR2006 2, which we did not observe (see Table \ref{sio_table_detections}). In addition, we made an additional detection, toward FMR2006 3. Our LSR velocities compare well with the values determined by \citet{nakashima06}.

We detected SiO maser emission  toward X 18 with a velocity of 71 \kms\ (see Table \ref{sio_table_detections}), which makes its cluster membership highly unlikely.  This value is in agreement with that measured at 43 GHz, i.e.,  30 \kms\ below the average cluster velocity. Were the maser star in a binary system, its velocity  could deviate from the average cluster velocity. However, the agreement between the two radial velocities taken at two different epochs appears to discard this hypothesis. Nevertheless, it is tempting to speculate that X 18 is a higher energy version of the famous R Aqr symbiotic binary system. This system consists of an M-type AGB star and a compact object, most likely a white dwarf. It hosts a strong SiO maser \citep{zuckerman79} and shows radio and ultraviolet emission but, in contrast to X 18, is not very luminous in X rays \citep{viotti1987,kafatos1989,kellogg2007}.

\subsection{RSGC2}

The RSGC2 was detected by \citet{davies07} as an overdensity of bright infrared stars in GLIMPSE and MSX images at
$l = 26\overset{\circ}{.}2, b = 0\overset{\circ}{.}0$. Remarkably, a total of 26 cluster members are  
spectroscopically confirmed RSG stars \citep{davies07} and  have spectral types from K2 to M5. In order to host such a large number of RSGs
implies a   mass of $4 \times 10^4$ \Msun\ for this cluster. From the average radial velocity ($109.3\pm0.7$ \kms) a
kinematic distance of $5.83^{+1.91}_{-0.78}$ kpc is inferred. By assuming this distance, and using theoretical models from
\citet{meynet00}, the authors derived a likely cluster age of $17\pm3$ Myr (or $12\pm1$ when considering non-rotating
isochrones). Luminosities $\log(L/\textrm{\Lsun})$ of the RSGs range from 4.4 to 5.4, suggesting initial stellar masses
of  about 15 \Msun\ \citep{meynet00}. All 26 RSGs were included in our SiO maser search program. We used the identification
numbers given by \citet{davies07}. We also included star number one, which has been classified as a
candidate RSG star in the foreground of the cluster because of its brightness ($K_S$=2.9 mag) and radial
velocity $\sim 95$ \kms\ \citep{davies07}.

Recently, \citet{deguchi10} performed an SiO $v = 1$ and $v = 2$, $J =1-0$ maser search at 43 GHz in 18 objects of the RSGC2 cluster, resulting in five detections. Stephenson 2 DFK 1 and DFK 2 were detected only in the $v = 1$ transition, while DFK 22, DFK 49 and 2MASS J18385699-0606459 showed maser emission in both the $v = 1$ and $v = 2$ lines.

We detected three of the objects toward which \citet{deguchi10}  found maser emission in the 43 GHz $J=1-0$ lines: Stephenson 2 DFK 1, DFK 2 and DFK 49. The LSR velocities derived from the 43 GHz maser lines are within a few km~s$^{-1}$ of the values we determine for the $J=2-1$ line, 
see  Table \ref{rsgc1_vlsr}.
We note that star number 1, which has a significantly different velocity than the other two, is likely a field RSG, not associated with RSGC2 \citep{davies07}.

\subsection{RSGC3}

The RSGC3 was identified by \citet{clark09}. It contains 8   RSGs with spectral types
from K3 to M4, and 8 other candidate RSGs on the basis of photometric information.  The number of RSGs suggests a
cluster mass of about $2 \times 10^4$ \Msun\ - $4 \times 10^4$ \Msun. The values of $\log(L/\textrm{\Lsun})$ range from
4.5 and 4.8 when assuming a distance of 6 kpc, and suggest initial masses similar to those of the RSGs in RSGC2.

The target selection and IRAM observations presented here were carried out before completion of the analysis of
\citet{clark09}. The 8  spectroscopically confirmed RSGs and one
of the candidate RSG were  unsuccessfully searched for SiO masers.  For those stars, we use the  identification numbers given by \citet{clark09}. Our list includes also 2MASS J18452254-0322261, which is not listed among the stellar members by \citet{clark09}.

\subsection{NGC 7419}

Before the discovery of the  RSGCs, the NGC 7419 cluster was the cluster with the largest number of known RSGs
\citep{caron03}. The cluster is located at $l = 109\overset{\circ}{.}14, b =
+1\overset{\circ}{.}14$ at a  photometric distance of $2.3 \pm 0.3$ kpc \citep{beauchamp94,caron03}, and has an age of $25\pm 5$ Myr
\citep{subramaniam06}. Its four RSG members have spectral types from M2.5 to M7.5. We list these RSGs with the names 
provided in the SIMBAD database. The  RSG MY Cep has already been detected in several maser surveys. 
It shows SiO, OH, and water masers \citep{nakashima07,sivagnanam90,takaba01}.

We only found maser emission at 86 GHz toward the brightest member of the NGC7419 cluster, the M7.5 star MY Cep. SiO Maser emission at 43 GHz had been detected before by \citet{nakashima07}. They derived a radial  velocity of $-53.7$ \kms\ from the $v = 1, J=1-0$ transition and $-51.9$ \kms\ from the $v=2, J=1-0$ line. Our observations resulted in a very similar LSR velocity of $-52$ \kms.
(see Table \ref{rsgc1_vlsr}).

\subsection{Perseus OB1}\label{section_perob1}

The  Perseus OB1 association contains the famous double cluster h/$\chi$ Per. These young clusters (age $\sim$14 My) have distances of $2.34^{+0.08}_{-0.08}$ kpc and $2.29^{+0.09}_{-0.08}$ kpc, respectively \citep{curie10}. The association
contains at least 22 RSGs \citep{Humphreys1978,garmany92,pierce00}. For one of them, S Per, recently a trigonometric parallax distance of $2.42^+{0.11}_{-0.09}$~kpc was measured \citep{asaki10}; see \citet{Xu2006} for another VLBI distance and a relevant discussion.

No SiO maser survey of the complete sample of RSGs in Per OB1 had been conducted so far. Only a few sources had been searched for SiO masers and maser emission was detected before in W Per \citep{jiang96} and the well studied S Per, which is known for its strong maser emission \citep[e.g][]{kaifu75}. The $^{28}$SiO $v = 1$ and $v = 2$, $J =1-0$ and $^{29}$SiO $v = 0$, $J =1-0$ masers were searched for toward FZ Per and T Per, but no emission was detected \citep{jiang99}. SU Per was targeted in a SiO $v = 1$, $v = 2$ and $v = 3$, $J =1-0$ search, but no maser was found \citep{spencer81}.

W Per was found to show emission in the 43 GHz $v = 1$, $J =1-0$ transition \citep{jiang96}. S Per was observed by various authors in a number of transitions. The first detection was the $v=1,J=2-1$ line at 86 GHz by \citet{kaifu75}. Furthermore, maser emission was found in the SiO  $v = 1$, $v = 2$ and $v = 3$, $J =1-0$ transitions \citep[e.g.][]{spencer77,cho96}, the $v = 0$ and $v = 2$, $J =2-1$  transitions \citep[e.g][]{morris79,olofsson81}, and the $v = 1$, $J=3-2$ transition \citep{herpin98}. \citet{pardo98} studied more than 15 different transitions of the SiO maser in S Per, detecting some high transitions such as $v = 2$ and $v = 3$, $J = 4-3$ and $v = 1$, $J=5-4$.

The RSGs in the Perseus OB1 association discussed by
\citet{pierce00} were part of our target list.  We also observed the brightest member of the NGC 457 cluster, which is a M1.5 Iab star \citep{mermilliod08}.

We detected SiO maser emission toward two RSGs in the Per OB1 association, W Per and S Per. Both sources were known to host masers (see above). The detected stars are among the three RSG members with longer periods and have redder colors of 485 and 822 days,  respectively. This led \citet{pierce00} to suggest that   these objects are surrounded by a CSE.

We obtain an LSR velocity of $-41$ \kms\ for S Per and $-40$ \kms\ for W Per, respectively (see Table \ref{sio_table_detections}). \citet{jiang96} reported  $-41.3$ \kms\ for W Per based on their 43 GHz maser observations, in agreement with our findings. For S Per, several results can be found in the literature that agree within a km~s$^{-1}$ with our value \citep{alcolea90,kim10}.

\subsection{GLIMPSE13}

The GLIMPSE13 cluster contains a large number of spectroscopically  confirmed
G9 and M1  stars  \citep{messineo09, mercer05}. including suspected supergiants with spectral types  ranging from
early K to late G. Since yellow supergiants are  rarer than RSGs, and Galactic RSGs have a median
spectral type of M2, it is more likely that the stars in GLIMPSE13 are massive giants
 with an age between 50 and 100 Myr \citep{messineo09}.
The bright star GLIMPSE13 1, however, is too luminous to be a giant member of the same cluster ($K_S$=2.67 mag) and is of type M0I or M7III.

GLIMPSE13 was targeted in the searches for SiO and water masers in stellar clusters conducted by 
\citet{deguchi10}. There is only an overlap of two sources with the stars for which \citet{messineo09} report IR data. \citet{deguchi10}
detected three SiO masing stars (including  GLIMPSE13 1, our one target star) with radial velocities around 70 \kms\  (72.3, 74.9 and
70.2 \kms), and they found water maser emission with a velocity of 21.4 \kms\ toward  GLIMPSE13 2 (a K5III star).  A
velocity of 71 \kms\ corresponds to a near-kinematic distance of $4.3_{-0.36}^{+0.37}$ kpc, while a velocity of 21
\kms\ implies a distance of $1.6\pm0.4$ kpc. The cluster is located in a particularly rich portion of  the disk at a longitude of
33.8$^\circ$. Several young stellar objects are located a few
arcminutes north of the cluster, and a number of bright mid-infrared sources surround the cluster. It is most
likely that we are seeing  different groups of stars along neighboring lines-of-sight that are not physically associated.

We detected SiO maser emission in the $v = 1$, $J=2-1$ line in the RSG candidate GLIMPSE13 1. Maser emission at 43 GHz had been observed before by \citet{deguchi10} in the $v = 1$, $J=1-0$ transition. They found  LSR  velocities of peak emission of 69.4 \kms\ and 72.3 \kms\ on two separate observing days, completely in agreement with our value of 71 \kms\ (see Table \ref{sio_table_detections}). The SiO maser lines' shape and breadth suggest that GLIMPSE13 1 is a RSG. The mean equivalent width of the maser emission is 27 \kms.

\subsection{GLIMPSE9}

The stellar cluster number 9 in the list of \citet{mercer05} was analyzed with near-infrared spectroscopy and HST/NICMOS photometry by
\citet{messineo09}. These authors detected 3 RSG stars and one late M giant (or early type RSG).  The cluster is located at a spectro-photometric distance
of 4.2 kpc, has a likely age of 15-27 Myr, and a cluster mass of at least 1600 \Msun.

We included in our target list all 4 late-type stars detected in the direction of the GLIMPSE9 cluster (3 RSGs and 1 giant). Maser lines were not detected.

\subsection{SGR 1900+14}

We also included among the targets the two RSG stars that are members of the stellar cluster associated with the magnetar (soft gamma ray repeater) SGR1900+14
\citep{davies10}. The cluster is located at a  kinematic distance of $12.5\pm1.7$ kpc, and has a likely age of 14 Myr. The two M5I stars have an
angular separation of just $\sim 2''$. Therefore only a single pointing was needed with the IRAM 30m telescope. Maser lines were not detected.

\subsection{MFD2008 (Cl 1813-178)}

A young massive cluster was recently identified in the direction of the W33 complex at a  kinematic distance of about 4 kpc. It contains several
evolved early-type stars and 1 RSG \citep{messineo08,messineo11w33}, from which a likely age of 4.5 Myr and a cluster mass of 10$^4$
\Msun\ are inferred. We included among our targets the RSG identified in the MFD2008 cluster by \citet{messineo08}. We did not detect SiO maser emission toward this cluster member.

\subsection{Trumpler 27}

The Trumpler 27 cluster is located at $l = 335\overset{\circ}{.}1, b = -0\overset{\circ}{.}7$ at a  spectro-photometric distance of $2.1 \pm 0.2$ kpc 
\citep{moffat77}. It contains a binary RSG  of M0 type \citep{massey01}. The presence of two Wolf Rayet stars and several other 
evolved OB stars suggests a cluster age below 6 Myr. We included in our program the RSG. The 43 GHz $v = 1$, $J=1-0$ transition was already observed by \citet{hall90}. The
86 GHz $v = 1$, $J=2-1$ line was also found before \citep{haikala94}.

We detected SiO maser emission at 86 GHz toward the binary RSG in Trumpler 27.  We determine an LSR velocity of $-17$ \kms\ (see Table \ref{sio_table_detections}), slightly higher than the $-11$ \kms\ reported by \citet{haikala94}.

\subsection{\label{isolated}The Galactic center and Inner Galaxy RSGs not associated with clusters}

\citet{blum03} performed a spectro-photometric analysis of a sample of 79 luminous late-type stars in the central parsecs of the
Galaxy. We observed 7 out of the 17 RSGs listed by \citet{blum03}, but no maser lines were detected.  One of the sources, BSD96 66, is the famous IRS 7, the brightest near IR source in the central parsec \citep{Becklin1975}. Toward this source, \citet{Menten1997} detected the 43.3 GHz $v = 1$, $J=1-0$ SiO line with the NRAO Very Large Array (VLA) with a peak  flux density of 0.38 Jy. Given that the 43 and 86 GHz
$v = 1$ maser fluxes are comparable (see Sect. \ref{intensities}), it is not surprising that we did not obtain a detection given our noise level (0.22 Jy; see Table \ref{sio_table_nondetections}).

With a similar technique, \citet{comeron04} identified  18 RSGs located in  the Inner Galaxy. We observed nine of these RSGs, but maser emission was not detected.

\subsection{Clusters without SiO maser detections -- summary}

With our average rms noise of 0.23 Jy, we report the first unsuccessful searches toward  targets in RSGC3, GLIMPSE9, MFD2008 and SGR 1900+14, as well as the isolated RSGs \citep{blum03,comeron04}  discussed in Sect. \ref{isolated} (see Table
\ref{sio_table_nondetections}). SiO masers vary with  pulsation phase \citep[e.g.][]{alcolea90,alcolea99}. New and deeper SiO maser observations could be useful to detect possible maser emission.

\section{Analysis\label{analysis}}

\subsection{Line equivalent widths}

\begin{figure}[h!]
\begin{center}
\includegraphics[angle=180,width=0.4\textwidth]{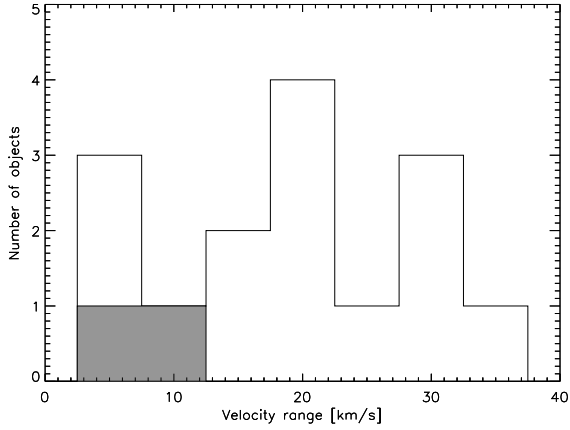}
\caption{Histogram of the LSR velocity range covered by the maser emission. The grey region shows the Mira variables IK Tau and U Her.}
\label{histo_vel_range}
\end{center}
\end{figure}

For comparison, we also observed the maser emission from the Mira variables IK Tau and U Her. From Table \ref{sio_table_detections} and Fig.
\ref{sio_detections_spectra}, it is obvious that the emission from the Miras, while very strong because close to Earth ($\sim 100$ pc),
covers much smaller velocity intervals than the RSGs' emission. The  SiO maser lines in RSGs are generally broader than those
in Mira stars \citep[e.g.][]{lebertre90}; within errors our new data confirms this previous result). Fig. \ref{histo_vel_range} shows a histogram of the velocity range covered by the SiO maser line. The Mira
variables IK Tau and U Her are shown in a grey color. Clearly, their profiles are much narrower than those of the bulk of the  RSG's. \citet{alcolea90} found the mean
equivalent width, given by (profile-area)/(peak-intensity), of the 86 GHz SiO maser lines to be 8.6 \kms\ (rms of 1.3 \kms) for their sample of 10 RSGs, and only 4 \kms\ (rms of 1 \kms)
for the 17 Miras. In our case, using the values of Table \ref{sio_table_detections}, the 13 RSGs have a mean equivalent width of 7.4 \kms (rms of 2.7 \kms), while the 2 Miras
have a width of 2.4 \kms (rms of 0.4 \kms). 

 Since for a CSE created by mass outflow, the terminal expansion velocity is proportional to the mass loss rate, the higher line widths we find for the RSGs (compared to Miras), simply reflects their greater degree of mass loss.

\subsection{Maser occurrence versus the Q1 and Q2 parameters}
With a combination of near- and mid-IR colors it is possible to identify 
mass-losing
stars and to distinguish between highly reddened ``naked" stars and
dust-enshrouded stars. \citet{messineo11}  proposed a new source 
selection method for identifying mass-losing
late-type stars, which is based on photometric data from 2MASS and GLIMPSE
\citep{cutri03,benjamin03} and on the  $Q1$ and $Q2$ parameters. These 
two parameters are
independent of interstellar extinction. The $Q1$ parameter 
($Q1=(J-H)-1.8 \times (H-K_S$)) is a
measure of the deviation from the reddening vector in the 2MASS $(J-H)$ 
versus
$(H-K_S)$ plane ($H, J,$ and $K_S$ are the stars' magnitudes in the corresponding IR bands).
Early-type stars have $Q1$ values  around 1 and
K-giants
around 0.4 mag. Dusty circumstellar envelopes change the  energy 
distribution of late-type stars,
and generate smaller $Q1$ values.  On average,  higher  mass-loss rates 
produce smaller  $Q1$
values \citep[see trend in Fig. 4 by][]{messineo11}.
Similarly, the $Q2$ parameter ($Q2=(J-K_S)-2.69\times(K_S-[8.0])$)
measures the deviation from the interstellar vector in the 2MASS/GLIMPSE 
plane $(J-K_S)$ versus
$(K_S-[8.0])$.  On average, higher mass-loss
rates imply smaller $Q2$ values \citep[see trends in Figs. 2 and 6 
by][]{messineo11}.

\begin{figure}[h!]
\begin{center}
\includegraphics[width=0.42\textwidth,angle=0]{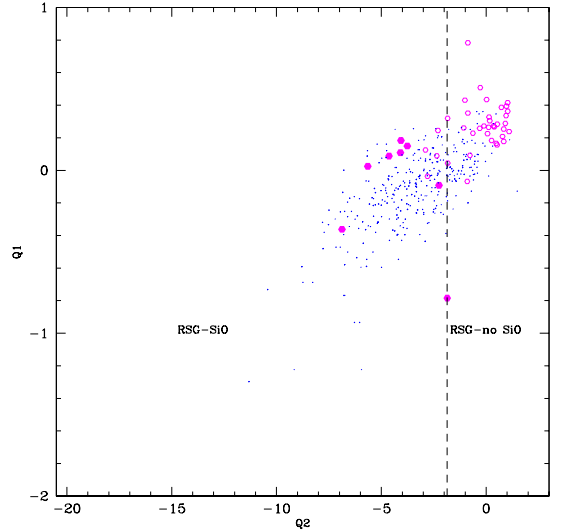}
\caption{\label{sio_colors}  The two extinction-free parameters Q1 and Q2 are plotted. Small dots
indicate SiO maser detections by \citet{messineo02}. Empty hexagons are our targeted RSGs without maser 
detections, and filled hexagons are those with maser emission.  }
\end{center}
\end{figure}

A plot of the $Q1$ versus $Q2$ parameters for the targeted RSGs is shown in Fig.
\ref{sio_colors}. It shows that with our sensitivity we were able to detect masers only in stars
with $Q2$ values smaller than
$-1.9$ mag. This implies that SiO maser emission was detected toward the reddest
sources,
and  supports a radiative pumping mechanism for the  SiO masers in
RSGs \citep{alcolea90}. For comparison, points representing the SiO maser star sample of
\citet{messineo02} are also plotted in Fig. \ref{sio_colors}.
The corresponding data were taken with the same telescope as for our sample  and have similar rms noise.
However, the Messineo et al.
sample is composed of Mira-like AGB stars, i.e. of stars  of luminosity class III,
\citep{messineo04,messineo05}, while our new data points represent RSGs
(luminosity class I).

There is an interesting  shift between the $Q2$ values of  masing sources from the one to the other
sample.
This suggests that higher mass-loss rates are required for RSGs to show SiO maser action than for AGB stars.
Perhaps, the
higher luminosities of RSGs push the dust condensation radius and the masing shell
farther out. Mass-loss is
indeed proportional to density and envelope size \citep[see equation A3
in][]{sevenster99}. A large statistics
is needed to confirm this suggested color threshold.

\begin{table*}[t]
\begin{center}
\caption{\label{tabratios}Maser/maser and maser/8 $\mu$m IR continuum flux ratios}
\begin{tabular}{lccc}
 \hline \hline
Object  & F$_{\rm 43 GHz}$/F$_{\rm 86 GHz}$  & F$_{\rm 86 GHz}$/F$_{\rm 8 \mu m}$  & Ref. 43 GHz  \\
\hline 
FMR2006 1                       & 0.76  & \dots &  (1) \\ 
FMR2006 3                       & \dots & \dots  &      \\ 
FMR2006 4                       & 0.89  & \dots &  (1) \\ 
FMR2006 13                     & 0.89  & 0.18  &  (1) \\
X 18                                    & 0.95  & 0.90  &  (1) \\
Stephenson 2 DFK 2       & 1.26  & 0.08  &  (2) \\
Stephenson 2 DFK 49     & 0.95  & 0.82  &  (2) \\
Stephenson 2 DFK 1       & 0.35  & 0.10  &  (2) \\
NGC 7419/MY Cep          & 1.38  & 0.09  &  (3) \\ 
W Per                                  & 0.41  & \dots &  (4) \\
S Per                                   & 0.21  & \dots &  (5) \\
MDI2009 GLIMPSE13 1  & 0.41  & 0.07  &  (2) \\
Cl Trumpler 27 1               & 0.16  & 0.33  &  (6) \\
IK Tau (Mira)                      & 1.22  & \dots &  (7) \\
U Her  (Mira)                      & 0.37  & \dots &  (7) \\
\noalign{\smallskip}
\hline
\noalign{\smallskip}
\end{tabular}
\end{center}
\small
\textbf{Notes.} Columns are (from left to right) designations of the targeted RSGs ratios of SiO(43 GHz peak flux)/SiO(86 GHz peak flux) ratio and SiO(86 GHz peak flux)/IR(8 $\mu$m flux) \citep[see][]{alcolea90}. The 8 $\mu$m fluxes are taken from the MSX and GLIMPSE catalogue. Sources without measurements were not included in these surveys, or saturated in GLIMPSE. References are:\\
\\
\textbf{References.} (1) \citet{nakashima06}; (2) \citet{deguchi10}; (3) \citet{nakashima07}; (4) \citet{jiang96}; (5) \citet{cho96};
(6) \citet{hall90}; (7) \citet{kim10}.
\end{table*}

\subsection{\label{intensities}SiO maser intensities}

In Table \ref{tabratios}, we list the ratio of the 43 GHz peak flux from the literature over the 86 GHz peak flux derived here. Apart from IK Tau, MY Cep and Cl*
Stephenson 2 DFK 2, the 86 GHz transition is found to be stronger in all sources. The mean value of the 43/86 GHz ratio is 0.73 with a standard deviation of 0.41. This makes the 86 GHz transition a good line for determining radial velocities of RSG stars. \citet{nyman93} concluded that the 43 GHz over 86 GHz ratio increases with increasing mass-loss rates.
Values above average are measured for MY  Cep and DFK 2. We should note, however,
that the SiO maser intensity is found to vary during the stellar variability cycle \citep[e.g.][]{alcolea90,alcolea99}. The 43 GHz and 86 GHz observations are not simultaneously taken, and their peak intensities could vary
with stellar phase. Therefore, the individual ratios may be not conclusive, and a simultaneous detection program
should be carried out. A large fraction of stars have ratios below unity, confirming that with single dish telescopes surveys for 86 GHz SiO maser emission are more efficient than for 43 GHz emission. 
One should however, note that the VLA has been a very sensitive instrument for 43 GHz SiO maser searches \citep[see ][]{Menten1997} and its expanded version, the Karl G. Jansky Very Large Array (JLVA) will be even more so. However, its sensitivity will be matched by the Atacama Large Millimeter Array (ALMA), which (after completion) will cover both lines with formidable sensitivity for sources with declinations up to $\sim 40^\circ$.

We also calculated the ratio of the 86 GHz peak flux over the IR 8 $\mu$m peak flux (see Table \ref{tabratios}). The 8 $\mu$m fluxes were taken from the MSX
\citep{price01} and GLIMPSE \citep{benjamin03} catalogues when available. Previously, \citet{alcolea90} performed a SiO maser search in Miras and supergiants, and
found a SiO(86 GHz peak flux)/IR(8$\mu$m flux) ratio $\sim$ 1/5 for Miras and the RSGs with the strongest maser emission, but significantly lower for the other supergiants (with 0.003 for
SW Vir being the lowest ratio). We find a mean value of 0.32 for the 86 GHz peak flux over the 8 $\mu$m flux ratio (standard deviation of 0.34). 
Clearly, for all stars, the SiO maser flux is smaller than the $8~\mu$m flux. Therefore they meet the criterion required by a radiative pump, which was established by the modeling of SiO maser pumping schemes; see Sect. \ref{maserpumping}. Again, we should note that both SiO maser and IR flux, which both are variable, generally were not measured contemporaneously.

\section{Conclusions\label{conclusions}}

We performed an extensive search for the SiO $v=1$, $J=2-1$ maser line at 86 GHz in 88 (candidate) RSGs in Galactic open clusters. Most of these stellar clusters were never before covered covered by  a 86 GHz SiO maser survey. In total, we detected maser emission from 13 RSGs, 1 of which is the first 
detection on any SiO in the respective object. The
measurements provide accurate LSR velocities, which can be used in studies of the Galactic structure. For the clusters containing masing stars, our non-detection of SiO emission in other RSG members may, in combination with a well determined distance to the cluster, yield information on the fraction of RSGs with high enough mass-loss to form maser containing CSEs or a dependence on maser occurrence with spectral type.

The FWHMs of  the SiO maser lines are broader in  RSGs than in Mira
AGBs. The intensities of the 86 GHz SiO maser lines are somewhat stronger than those of 43 GHz  $v=1$, $J=1-0$ SiO maser lines in  most the sources (11 out of the 14 known to be masing in both lines). The JVLA and ALMA will provide all sky coverage for extremely sensitive surveys in both lines.

SiO maser emission was detected in targets with $Q2$ values larger than $\sim -2$ mag. This increase of the detection rate with redder colors supports a radiative pumping mechanism for the SiO maser. The $Q1$ and $Q2$ parameters are proven to be useful tools for efficient targeted searches for SiO masers.

\begin{acknowledgements}
We thank the IRAM 30m staff for the support during the observations. LV was supported for this research through a stipend from the International Max Planck Research School (IMPRS) for Astronomy and Astrophysics at the Universities of Bonn and Cologne. MM thanks  Don Figer, Ben Davies, Simon Clark, and Ignatio Negueruela, for creative discussions on red supergiant clusters (RSGCs). The co-proposed SiO observations (non-detections) on RSGC3  are here included. This research has made use of the SIMBAD database, operated at CDS, Strasbourg, France and was partially funded by the ERC Advanced Investigator Grant GLOSTAR (247078).
\end{acknowledgements}

\bibliographystyle{aa}
\bibliography{18265.bib}

\begin{thebibliography}{98}
\expandafter\ifx\csname natexlab\endcsname\relax\def\natexlab#1{#1}\fi

\bibitem[{{Alcolea} {et~al.}(1990){Alcolea}, {Bujarrabal}, \&
  {Gomez-Gonzalez}}]{alcolea90}
{Alcolea}, J., {Bujarrabal}, V., \& {Gomez-Gonzalez}, J. 1990, \aap, 231, 431

\bibitem[{{Alcolea} {et~al.}(1999){Alcolea}, {Pardo}, {Bujarrabal},
  {Bachiller}, {Barcia}, {Colomer}, {Gallego}, {G{\'o}mez-Gonz{\'a}lez}, {del
  Pino Cisneros}, {Planesas}, {del R{\'{\i}}o}, {Rodr{\'{\i}}guez-Franco}, {del
  Romero}, {Tafalla}, \& {de Vicente}}]{alcolea99}
{Alcolea}, J., {Pardo}, J.~R., {Bujarrabal}, V., {et~al.} 1999, \aaps, 139, 461

\bibitem[{{Alexander} {et~al.}(2009){Alexander}, {Kobulnicky}, {Clemens},
  {Jameson}, {Pinnick}, \& {Pavel}}]{alexander09}
{Alexander}, M.~J., {Kobulnicky}, H.~A., {Clemens}, D.~P., {et~al.} 2009, \aj,
  137, 4824

\bibitem[{{Asaki} {et~al.}(2010){Asaki}, {Deguchi}, {Imai}, {Hachisuka},
  {Miyoshi}, \& {Honma}}]{asaki10}
{Asaki}, Y., {Deguchi}, S., {Imai}, H., {et~al.} 2010, \apj, 721, 267

\bibitem[{{Beauchamp} {et~al.}(1994){Beauchamp}, {Moffat}, \&
  {Drissen}}]{beauchamp94}
{Beauchamp}, A., {Moffat}, A.~F.~J., \& {Drissen}, L. 1994, \apjs, 93, 187

\bibitem[{{Becklin} \& {Neugebauer}(1975)}]{Becklin1975}
{Becklin}, E.~E. \& {Neugebauer}, G. 1975, \apjl, 200, L71

\bibitem[{{Benjamin} {et~al.}(2003){Benjamin}, {Churchwell}, {Babler}, {Bania},
  {Clemens}, {Cohen}, {Dickey}, {Indebetouw}, {Jackson}, {Kobulnicky},
  {Lazarian}, {Marston}, {Mathis}, {Meade}, {Seager}, {Stolovy}, {Watson},
  {Whitney}, {Wolff}, \& {Wolfire}}]{benjamin03}
{Benjamin}, R.~A., {Churchwell}, E., {Babler}, B.~L., {et~al.} 2003, \pasp,
  115, 953

\bibitem[{{Benson} {et~al.}(1990){Benson}, {Little-Marenin}, {Woods},
  {Attridge}, {Blais}, {Rudolph}, {Rubiera}, \& {Keefe}}]{besanson90}
{Benson}, P.~J., {Little-Marenin}, I.~R., {Woods}, T.~C., {et~al.} 1990, \apjs,
  74, 911

\bibitem[{{Bissantz} {et~al.}(2003){Bissantz}, {Englmaier}, \&
  {Gerhard}}]{bissantz03}
{Bissantz}, N., {Englmaier}, P., \& {Gerhard}, O. 2003, \mnras, 340, 949

\bibitem[{{Blum} {et~al.}(2003){Blum}, {Ram{\'{\i}}rez}, {Sellgren}, \&
  {Olsen}}]{blum03}
{Blum}, R.~D., {Ram{\'{\i}}rez}, S.~V., {Sellgren}, K., \& {Olsen}, K. 2003,
  \apj, 597, 323

\bibitem[{{Buhl} {et~al.}(1974){Buhl}, {Snyder}, {Lovas}, \&
  {Johnson}}]{buhl74}
{Buhl}, D., {Snyder}, L.~E., {Lovas}, F.~J., \& {Johnson}, D.~R. 1974, \apjl,
  192, L97

\bibitem[{{Bujarrabal}(1994{\natexlab{a}})}]{bujarrabal94a}
{Bujarrabal}, V. 1994{\natexlab{a}}, \aap, 285, 953

\bibitem[{{Bujarrabal}(1994{\natexlab{b}})}]{bujarrabal94b}
{Bujarrabal}, V. 1994{\natexlab{b}}, \aap, 285, 971

\bibitem[{{Bujarrabal} {et~al.}(1996){Bujarrabal}, {Alcolea}, {Sanchez
  Contreras}, \& {Colomer}}]{bujarrabal96}
{Bujarrabal}, V., {Alcolea}, J., {Sanchez Contreras}, C., \& {Colomer}, F.
  1996, \aap, 314, 883

\bibitem[{{Bujarrabal} {et~al.}(1987){Bujarrabal}, {Planesas}, \& {del
  Romero}}]{bujarrabal87}
{Bujarrabal}, V., {Planesas}, P., \& {del Romero}, A. 1987, \aap, 175, 164

\bibitem[{{Caron} {et~al.}(2003){Caron}, {Moffat}, {St-Louis}, {Wade}, \&
  {Lester}}]{caron03}
{Caron}, G., {Moffat}, A.~F.~J., {St-Louis}, N., {Wade}, G.~A., \& {Lester},
  J.~B. 2003, \aj, 126, 1415

\bibitem[{{Cho} {et~al.}(1996){Cho}, {Kaifu}, \& {Ukita}}]{cho96}
{Cho}, S.-H., {Kaifu}, N., \& {Ukita}, N. 1996, \aaps, 115, 117

\bibitem[{{Choi} {et~al.}(2008){Choi}, {Hirota}, {Honma}, {Kobayashi},
  {Bushimata}, {Imai}, {Iwadate}, {Jike}, {Kameno}, {Kameya}, {Kamohara},
  {Kan-Ya}, {Kawaguchi}, {Kijima}, {Kim}, {Kuji}, {Kurayama}, {Manabe},
  {Maruyama}, {Matsui}, {Matsumoto}, {Miyaji}, {Nagayama}, {Nakagawa},
  {Nakamura}, {Oh}, {Omodaka}, {Oyama}, {Sakai}, {Sasao}, {Sato}, {Sato},
  {Shibata}, {Tamura}, {Tsushima}, \& {Yamashita}}]{choi08}
{Choi}, Y.~K., {Hirota}, T., {Honma}, M., {et~al.} 2008, \pasj, 60, 1007

\bibitem[{{Clark} {et~al.}(2005){Clark}, {Negueruela}, {Crowther}, \&
  {Goodwin}}]{clark05}
{Clark}, J.~S., {Negueruela}, I., {Crowther}, P.~A., \& {Goodwin}, S.~P. 2005,
  \aap, 434, 949

\bibitem[{{Clark} {et~al.}(2009){Clark}, {Negueruela}, {Davies}, {Larionov},
  {Ritchie}, {Figer}, {Messineo}, {Crowther}, \& {Arkharov}}]{clark09}
{Clark}, J.~S., {Negueruela}, I., {Davies}, B., {et~al.} 2009, \aap, 498, 109

\bibitem[{{Comer{\'o}n} {et~al.}(2004){Comer{\'o}n}, {Torra}, {Chiappini},
  {Figueras}, {Ivanov}, \& {Ribas}}]{comeron04}
{Comer{\'o}n}, F., {Torra}, J., {Chiappini}, C., {et~al.} 2004, \aap, 425, 489

\bibitem[{{Cotera} {et~al.}(1996){Cotera}, {Erickson}, {Colgan}, {Simpson},
  {Allen}, \& {Burton}}]{cotera96}
{Cotera}, A.~S., {Erickson}, E.~F., {Colgan}, S.~W.~J., {et~al.} 1996, \apj,
  461, 750

\bibitem[{{Currie} {et~al.}(2010){Currie}, {Hernandez}, {Irwin}, {Kenyon},
  {Tokarz}, {Balog}, {Bragg}, {Berlind}, \& {Calkins}}]{curie10}
{Currie}, T., {Hernandez}, J., {Irwin}, J., {et~al.} 2010, \apjs, 186, 191

\bibitem[{{Cutri} {et~al.}(2003){Cutri}, {Skrutskie}, {van Dyk}, {Beichman},
  {Carpenter}, {Chester}, {Cambresy}, {Evans}, {Fowler}, {Gizis}, {Howard},
  {Huchra}, {Jarrett}, {Kopan}, {Kirkpatrick}, {Light}, {Marsh}, {McCallon},
  {Schneider}, {Stiening}, {Sykes}, {Weinberg}, {Wheaton}, {Wheelock}, \&
  {Zacarias}}]{cutri03}
{Cutri}, R.~M., {Skrutskie}, M.~F., {van Dyk}, S., {et~al.} 2003, {2MASS All
  Sky Catalog of point sources.} ({NASA})

\bibitem[{{Davies} {et~al.}(2007){Davies}, {Figer}, {Kudritzki}, {MacKenty},
  {Najarro}, \& {Herrero}}]{davies07}
{Davies}, B., {Figer}, D.~F., {Kudritzki}, R., {et~al.} 2007, \apj, 671, 781

\bibitem[{{Davies} {et~al.}(2009{\natexlab{a}}){Davies}, {Figer}, {Kudritzki},
  {Trombley}, {Kouveliotou}, \& {Wachter}}]{davies10}
{Davies}, B., {Figer}, D.~F., {Kudritzki}, R., {et~al.} 2009{\natexlab{a}},
  \apj, 707, 844

\bibitem[{{Davies} {et~al.}(2008){Davies}, {Figer}, {Law}, {Kudritzki},
  {Najarro}, {Herrero}, \& {MacKenty}}]{davies08}
{Davies}, B., {Figer}, D.~F., {Law}, C.~J., {et~al.} 2008, \apj, 676, 1016

\bibitem[{{Davies} {et~al.}(2009{\natexlab{b}}){Davies}, {Origlia},
  {Kudritzki}, {Figer}, {Rich}, {Najarro}, {Negueruela}, \& {Clark}}]{davies09}
{Davies}, B., {Origlia}, L., {Kudritzki}, R., {et~al.} 2009{\natexlab{b}},
  \apj, 696, 2014

\bibitem[{{Deguchi} {et~al.}(2002){Deguchi}, {Fujii}, {Nakashima}, \&
  {Wood}}]{Deguchi2002}
{Deguchi}, S., {Fujii}, T., {Nakashima}, J.-I., \& {Wood}, P.~R. 2002, \pasj,
  54, 719

\bibitem[{{Deguchi} {et~al.}(2010){Deguchi}, {Nakashima}, {Zhang}, {Chong},
  {Koike}, \& {Kwok}}]{deguchi10}
{Deguchi}, S., {Nakashima}, J., {Zhang}, Y., {et~al.} 2010, \pasj, 62, 391

\bibitem[{{Elitzur}(1980)}]{elitzur80}
{Elitzur}, M. 1980, \apj, 240, 553

\bibitem[{{Epchtein} {et~al.}(1999){Epchtein}, {Deul}, {Derriere},
  {Borsenberger}, {Egret}, {Simon}, {Alard}, {Balazs}, {de Batz}, {Cioni},
  {Copet}, {Dennefeld}, {Forveille}, {Fouque}, {Garzon}, {Habing}, {Holl},
  {Hron}, {Kimeswenger}, {Lacombe}, {Le Bertre}, {Loup}, {Mamon}, {Omont},
  {Paturel}, {Persi}, {Robin}, {Rouan}, {Tiphene}, {Vauglin}, \&
  {Wagner}}]{denis2}
{Epchtein}, N., {Deul}, E., {Derriere}, S., {et~al.} 1999, VizieR Online Data
  Catalog, 1, 2002

\bibitem[{{Figer} {et~al.}(1999){Figer}, {Kim}, {Morris}, {Serabyn}, {Rich}, \&
  {McLean}}]{figer99}
{Figer}, D.~F., {Kim}, S.~S., {Morris}, M., {et~al.} 1999, \apj, 525, 750

\bibitem[{{Figer} {et~al.}(2006){Figer}, {MacKenty}, {Robberto}, {Smith},
  {Najarro}, {Kudritzki}, \& {Herrero}}]{figer06}
{Figer}, D.~F., {MacKenty}, J.~W., {Robberto}, M., {et~al.} 2006, \apj, 643,
  1166

\bibitem[{{Figer} {et~al.}(2002){Figer}, {Najarro}, {Gilmore}, {Morris}, {Kim},
  {Serabyn}, {McLean}, {Gilbert}, {Graham}, {Larkin}, {Levenson}, \&
  {Teplitz}}]{figer02}
{Figer}, D.~F., {Najarro}, F., {Gilmore}, D., {et~al.} 2002, \apj, 581, 258

\bibitem[{{Figer} {et~al.}(2004){Figer}, {Rich}, {Kim}, {Morris}, \&
  {Serabyn}}]{figer04}
{Figer}, D.~F., {Rich}, R.~M., {Kim}, S.~S., {Morris}, M., \& {Serabyn}, E.
  2004, \apj, 601, 319

\bibitem[{{Garmany} \& {Stencel}(1992)}]{garmany92}
{Garmany}, C.~D. \& {Stencel}, R.~E. 1992, \aaps, 94, 211

\bibitem[{{Habing}(1996)}]{habing96}
{Habing}, H.~J. 1996, \aapr, 7, 97

\bibitem[{{Habing} {et~al.}(2006){Habing}, {Sevenster}, {Messineo}, {van de
  Ven}, \& {Kuijken}}]{habing06}
{Habing}, H.~J., {Sevenster}, M.~N., {Messineo}, M., {van de Ven}, G., \&
  {Kuijken}, K. 2006, \aap, 458, 151

\bibitem[{{Haikala} {et~al.}(1994){Haikala}, {Nyman}, \&
  {Forsstroem}}]{haikala94}
{Haikala}, L.~K., {Nyman}, L., \& {Forsstroem}, V. 1994, \aaps, 103, 107

\bibitem[{{Hall} {et~al.}(1990){Hall}, {Allen}, {Troup}, {Wark}, \&
  {Wright}}]{hall90}
{Hall}, P.~J., {Allen}, D.~A., {Troup}, E.~R., {Wark}, R.~M., \& {Wright},
  A.~E. 1990, \mnras, 243, 480

\bibitem[{{Harwit} {et~al.}(2001){Harwit}, {Malfait}, {Decin}, {Waelkens},
  {Feuchtgruber}, \& {Melnick}}]{harwit01}
{Harwit}, M., {Malfait}, K., {Decin}, L., {et~al.} 2001, \apj, 557, 844

\bibitem[{{Herpin} {et~al.}(1998){Herpin}, {Baudry}, {Alcolea}, \&
  {Cernicharo}}]{herpin98}
{Herpin}, F., {Baudry}, A., {Alcolea}, J., \& {Cernicharo}, J. 1998, \aap, 334,
  1037

\bibitem[{{Heske}(1989)}]{heske89}
{Heske}, A. 1989, \aap, 208, 77

\bibitem[{{Humphreys}(2007)}]{Humphreys2007}
{Humphreys}, E.~M.~L. 2007, in IAU Symposium, Vol. 242, IAU Symposium, ed.
  {J.~M.~Chapman \& W.~A.~Baan}, 471--480

\bibitem[{{Humphreys} {et~al.}(2002){Humphreys}, {Gray}, {Yates}, {Field},
  {Bowen}, \& {Diamond}}]{humphreys02}
{Humphreys}, E.~M.~L., {Gray}, M.~D., {Yates}, J.~A., {et~al.} 2002, \aap, 386,
  256

\bibitem[{{Humphreys}(1978)}]{Humphreys1978}
{Humphreys}, R.~M. 1978, \apjs, 38, 309

\bibitem[{{Jewell} {et~al.}(1991){Jewell}, {Snyder}, {Walmsley}, {Wilson}, \&
  {Gensheimer}}]{jewell91}
{Jewell}, P.~R., {Snyder}, L.~E., {Walmsley}, C.~M., {Wilson}, T.~L., \&
  {Gensheimer}, P.~D. 1991, \aap, 242, 211

\bibitem[{{Jiang} {et~al.}(1999){Jiang}, {Deguchi}, \& {Ramesh}}]{jiang99}
{Jiang}, B.~W., {Deguchi}, S., \& {Ramesh}, B. 1999, \pasj, 51, 95

\bibitem[{{Jiang} {et~al.}(1996){Jiang}, {Deguchi}, {Yamamura}, {Nakada},
  {Cho}, \& {Yamagata}}]{jiang96}
{Jiang}, B.~W., {Deguchi}, S., {Yamamura}, I., {et~al.} 1996, \apjs, 106, 463

\bibitem[{{Kafatos} {et~al.}(1989){Kafatos}, {Hollis}, {Yusef-Zadeh},
  {Michalitsianos}, \& {Elitzur}}]{kafatos1989}
{Kafatos}, M., {Hollis}, J.~M., {Yusef-Zadeh}, F., {Michalitsianos}, A.~G., \&
  {Elitzur}, M. 1989, \apj, 346, 991

\bibitem[{{Kaifu} {et~al.}(1975){Kaifu}, {Buhl}, \& {Snyder}}]{kaifu75}
{Kaifu}, N., {Buhl}, D., \& {Snyder}, L.~E. 1975, \apj, 195, 359

\bibitem[{{Kellogg} {et~al.}(2007){Kellogg}, {Anderson}, {Korreck},
  {DePasquale}, {Nichols}, {Sokoloski}, {Krauss}, \& {Pedelty}}]{kellogg2007}
{Kellogg}, E., {Anderson}, C., {Korreck}, K., {et~al.} 2007, \apj, 664, 1079

\bibitem[{{Kim} {et~al.}(2010){Kim}, {Cho}, {Oh}, \& {Byun}}]{kim10}
{Kim}, J., {Cho}, S.-H., {Oh}, C.~S., \& {Byun}, D.-Y. 2010, \apjs, 188, 209

\bibitem[{{Kwan} \& {Scoville}(1974)}]{kwan74}
{Kwan}, J. \& {Scoville}, N. 1974, \apjl, 194, L97

\bibitem[{{Le Bertre} \& {Nyman}(1990)}]{lebertre90}
{Le Bertre}, T. \& {Nyman}, L. 1990, \aap, 233, 477

\bibitem[{{Levesque}(2010)}]{levesque10}
{Levesque}, E.~M. 2010, in Astronomical Society of the Pacific Conference
  Series, Vol. 425, Hot and Cool: Bridging Gaps in Massive Star Evolution, ed.
  {C.~Leitherer, P.~D.~Bennett, P.~W.~Morris, \& J.~T.~Van Loon}, 103

\bibitem[{{Lockett} \& {Elitzur}(1992)}]{lockett92}
{Lockett}, P. \& {Elitzur}, M. 1992, \apj, 399, 704

\bibitem[{{Massey} {et~al.}(2001){Massey}, {DeGioia-Eastwood}, \&
  {Waterhouse}}]{massey01}
{Massey}, P., {DeGioia-Eastwood}, K., \& {Waterhouse}, E. 2001, \aj, 121, 1050

\bibitem[{{Menten} {et~al.}(2008){Menten}, {Lundgren}, {Belloche}, {Thorwirth},
  \& {Reid}}]{Menten2008}
{Menten}, K.~M., {Lundgren}, A., {Belloche}, A., {Thorwirth}, S., \& {Reid},
  M.~J. 2008, \aap, 477, 185

\bibitem[{{Menten} {et~al.}(1997){Menten}, {Reid}, {Eckart}, \&
  {Genzel}}]{Menten1997}
{Menten}, K.~M., {Reid}, M.~J., {Eckart}, A., \& {Genzel}, R. 1997, \apjl, 475,
  L111

\bibitem[{{Mercer} {et~al.}(2005){Mercer}, {Clemens}, {Meade}, {Babler},
  {Indebetouw}, {Whitney}, {Watson}, {Wolfire}, {Wolff}, {Bania}, {Benjamin},
  {Cohen}, {Dickey}, {Jackson}, {Kobulnicky}, {Mathis}, {Stauffer}, {Stolovy},
  {Uzpen}, \& {Churchwell}}]{mercer05}
{Mercer}, E.~P., {Clemens}, D.~P., {Meade}, M.~R., {et~al.} 2005, \apj, 635,
  560

\bibitem[{{Mermilliod} {et~al.}(2008){Mermilliod}, {Mayor}, \&
  {Udry}}]{mermilliod08}
{Mermilliod}, J.~C., {Mayor}, M., \& {Udry}, S. 2008, \aap, 485, 303

\bibitem[{{Messineo} {et~al.}(2011){Messineo}, {Davies}, {Figer}, {Kudritzki},
  {Valenti}, {Trombley}, {Najarro}, \& {Rich}}]{messineo11w33}
{Messineo}, M., {Davies}, B., {Figer}, D.~F., {et~al.} 2011, \apj, 733, 41

\bibitem[{{Messineo} {et~al.}(2009){Messineo}, {Davies}, {Ivanov}, {Figer},
  {Schuller}, {Habing}, {Menten}, \& {Petr-Gotzens}}]{messineo09}
{Messineo}, M., {Davies}, B., {Ivanov}, V.~D., {et~al.} 2009, \apj, 697, 701

\bibitem[{{Messineo} {et~al.}(2008){Messineo}, {Figer}, {Davies}, {Rich},
  {Valenti}, \& {Kudritzki}}]{messineo08}
{Messineo}, M., {Figer}, D.~F., {Davies}, B., {et~al.} 2008, \apjl, 683, L155

\bibitem[{{Messineo} {et~al.}(2004){Messineo}, {Habing}, {Menten}, {Omont}, \&
  {Sjouwerman}}]{messineo04}
{Messineo}, M., {Habing}, H.~J., {Menten}, K.~M., {Omont}, A., \& {Sjouwerman},
  L.~O. 2004, \aap, 418, 103

\bibitem[{{Messineo} {et~al.}(2005){Messineo}, {Habing}, {Menten}, {Omont},
  {Sjouwerman}, \& {Bertoldi}}]{messineo05}
{Messineo}, M., {Habing}, H.~J., {Menten}, K.~M., {et~al.} 2005, \aap, 435, 575

\bibitem[{{Messineo} {et~al.}(2002){Messineo}, {Habing}, {Sjouwerman}, {Omont},
  \& {Menten}}]{messineo02}
{Messineo}, M., {Habing}, H.~J., {Sjouwerman}, L.~O., {Omont}, A., \& {Menten},
  K.~M. 2002, \aap, 393, 115

\bibitem[{{Messineo} {et~al.}(2012){Messineo}, {Menten}, {Churchwell}, \&
  {Habing}}]{messineo11}
{Messineo}, M., {Menten}, K.~M., {Churchwell}, E., \& {Habing}, H. 2012, \aap,
  537, A10

\bibitem[{{Meynet} \& {Maeder}(2000)}]{meynet00}
{Meynet}, G. \& {Maeder}, A. 2000, \aap, 361, 101

\bibitem[{{Moffat} {et~al.}(1977){Moffat}, {Fitzgerald}, \&
  {Jackson}}]{moffat77}
{Moffat}, A.~F.~J., {Fitzgerald}, M.~P., \& {Jackson}, P.~D. 1977, \apj, 215,
  106

\bibitem[{{Morris} {et~al.}(1979){Morris}, {Redman}, {Reid}, \&
  {Dickinson}}]{morris79}
{Morris}, M., {Redman}, R., {Reid}, M.~J., \& {Dickinson}, D.~F. 1979, \apj,
  229, 257

\bibitem[{{Nakashima} \& {Deguchi}(2006)}]{nakashima06}
{Nakashima}, J. \& {Deguchi}, S. 2006, \apjl, 647, L139

\bibitem[{{Nakashima} \& {Deguchi}(2007)}]{nakashima07}
{Nakashima}, J. \& {Deguchi}, S. 2007, \apj, 669, 446

\bibitem[{{Negueruela} {et~al.}(2011){Negueruela},
  {Gonz{\'a}lez-Fern{\'a}ndez}, {Marco}, \& {Clark}}]{negueruela11}
{Negueruela}, I., {Gonz{\'a}lez-Fern{\'a}ndez}, C., {Marco}, A., \& {Clark},
  J.~S. 2011, \aap, 528, A59+

\bibitem[{{Negueruela} {et~al.}(2010){Negueruela},
  {Gonz{\'a}lez-Fern{\'a}ndez}, {Marco}, {Clark}, \&
  {Mart{\'{\i}}nez-N{\'u}{\~n}ez}}]{negueruela10}
{Negueruela}, I., {Gonz{\'a}lez-Fern{\'a}ndez}, C., {Marco}, A., {Clark},
  J.~S., \& {Mart{\'{\i}}nez-N{\'u}{\~n}ez}, S. 2010, \aap, 513, A74+

\bibitem[{{Nguyen Luong} {et~al.}(2011){Nguyen Luong}, {Motte}, {Schuller},
  {Schneider}, {Bontemps}, {Schilke}, {Menten}, {Heitsch}, {Wyrowski},
  {Carlhoff}, {Bronfman}, \& {Henning}}]{Nguyen2011}
{Nguyen Luong}, Q., {Motte}, F., {Schuller}, F., {et~al.} 2011, \aap, 529, A41

\bibitem[{{Nyman} {et~al.}(1993){Nyman}, {Hall}, \& {Le Bertre}}]{nyman93}
{Nyman}, L.-A., {Hall}, P.~J., \& {Le Bertre}, T. 1993, \aap, 280, 551

\bibitem[{{Olofsson} {et~al.}(1981){Olofsson}, {Rydbeck}, {Lane}, \&
  {Predmore}}]{olofsson81}
{Olofsson}, H., {Rydbeck}, O.~E.~H., {Lane}, A.~P., \& {Predmore}, C.~R. 1981,
  \apjl, 247, L81

\bibitem[{{Pardo} {et~al.}(1998){Pardo}, {Cernicharo}, {Gonzalez-Alfonso}, \&
  {Bujarrabal}}]{pardo98}
{Pardo}, J.~R., {Cernicharo}, J., {Gonzalez-Alfonso}, E., \& {Bujarrabal}, V.
  1998, \aap, 329, 219

\bibitem[{{Pierce} {et~al.}(2000){Pierce}, {Jurcevic}, \&
  {Crabtree}}]{pierce00}
{Pierce}, M.~J., {Jurcevic}, J.~S., \& {Crabtree}, D. 2000, \mnras, 313, 271

\bibitem[{{Price} {et~al.}(2001){Price}, {Egan}, {Carey}, {Mizuno}, \&
  {Kuchar}}]{price01}
{Price}, S.~D., {Egan}, M.~P., {Carey}, S.~J., {Mizuno}, D.~R., \& {Kuchar},
  T.~A. 2001, \aj, 121, 2819

\bibitem[{{Reid} {et~al.}(2009){Reid}, {Menten}, {Zheng}, {Brunthaler},
  {Moscadelli}, {Xu}, {Zhang}, {Sato}, {Honma}, {Hirota}, {Hachisuka}, {Choi},
  {Moellenbrock}, \& {Bartkiewicz}}]{reid09}
{Reid}, M.~J., {Menten}, K.~M., {Zheng}, X.~W., {et~al.} 2009, \apj, 700, 137

\bibitem[{{Scholz} \& {Wood}(2000)}]{scholz00}
{Scholz}, M. \& {Wood}, P.~R. 2000, \aap, 362, 1065

\bibitem[{{Sevenster}(1999)}]{sevenster99}
{Sevenster}, M.~N. 1999, \mnras, 310, 629

\bibitem[{{Sivagnanam} {et~al.}(1990){Sivagnanam}, {Le Squeren}, {Minh}, \&
  {Braz}}]{sivagnanam90}
{Sivagnanam}, P., {Le Squeren}, A.~M., {Minh}, F.~T., \& {Braz}, M.~A. 1990,
  \aap, 233, 112

\bibitem[{{Spencer} {et~al.}(1977){Spencer}, {Schwartz}, {Waak}, \&
  {Bologna}}]{spencer77}
{Spencer}, J.~H., {Schwartz}, P.~R., {Waak}, J.~A., \& {Bologna}, J.~M. 1977,
  \aj, 82, 706

\bibitem[{{Spencer} {et~al.}(1981){Spencer}, {Schwartz}, {Winnberg}, {Olnon},
  {Matthews}, \& {Downes}}]{spencer81}
{Spencer}, J.~H., {Schwartz}, P.~R., {Winnberg}, A., {et~al.} 1981, \aj, 86,
  392

\bibitem[{{Stephenson}(1990)}]{stephenson90}
{Stephenson}, C.~B. 1990, \aj, 99, 1867

\bibitem[{{Subramaniam} {et~al.}(2006){Subramaniam}, {Mathew}, {Bhatt}, \&
  {Ramya}}]{subramaniam06}
{Subramaniam}, A., {Mathew}, B., {Bhatt}, B.~C., \& {Ramya}, S. 2006, \mnras,
  370, 743

\bibitem[{{Takaba} {et~al.}(2001){Takaba}, {Iwate}, {Miyaji}, \&
  {Deguchi}}]{takaba01}
{Takaba}, H., {Iwate}, T., {Miyaji}, T., \& {Deguchi}, S. 2001, \pasj, 53, 517

\bibitem[{{Valdettaro} {et~al.}(2001){Valdettaro}, {Palla}, {Brand},
  {Cesaroni}, {Comoretto}, {Di Franco}, {Felli}, {Natale}, {Palagi}, {Panella},
  \& {Tofani}}]{valdettaro01}
{Valdettaro}, R., {Palla}, F., {Brand}, J., {et~al.} 2001, \aap, 368, 845

\bibitem[{{Verhoelst} {et~al.}(2009){Verhoelst}, {van der Zypen}, {Hony},
  {Decin}, {Cami}, \& {Eriksson}}]{verhoelst09}
{Verhoelst}, T., {van der Zypen}, N., {Hony}, S., {et~al.} 2009, \aap, 498, 127

\bibitem[{{Viotti} {et~al.}(1987){Viotti}, {Piro}, {Friedjung}, \&
  {Cassatella}}]{viotti1987}
{Viotti}, R., {Piro}, L., {Friedjung}, M., \& {Cassatella}, A. 1987, \apjl,
  319, L7

\bibitem[{{Westerlund}(1961)}]{westerlund61}
{Westerlund}, B. 1961, \pasp, 73, 51

\bibitem[{{Xu} {et~al.}(2006){Xu}, {Reid}, {Zheng}, \& {Menten}}]{Xu2006}
{Xu}, Y., {Reid}, M.~J., {Zheng}, X.~W., \& {Menten}, K.~M. 2006, Science, 311,
  54

\bibitem[{{Zuckerman}(1979)}]{zuckerman79}
{Zuckerman}, B. 1979, \apj, 230, 442

\end{thebibliography}

\longtab{3}{
\begin{longtable}{lccc}
\caption{\label{sio_table_nondetections}Red Supergiants without SiO~$\textit{v} = 1; J = 2-1$ detection.} \\
 \hline \hline
Object  & $\alpha_{J2000}$  & $\delta_{J2000}$ & rms  \\
  &  [J2000]  & [J2000]  & [Jy] \\
\hline
\multicolumn{4}{c}{\citet[RSGC1 cluster]{figer06}}\\
FMR2006 11 & 18:37:51.72 & -06:51:49.9  & 0.497 \\
FMR2006 12 & 18:38:03.30 & -06:52:45.1  & 0.499 \\
FMR2006 14 & 18:37:47.64 & -06:53:02.3  & 0.567 \\
FMR2006 16 & 18:38:01.29 & -06:52:51.9  & 0.590 \\ 
\multicolumn{4}{c}{\citet[RSGC2 cluster]{davies07}}\\
Stephenson 2 DFK 3  & 18:39:24.60  & -06:02:13.8  & 0.186 \\
Stephenson 2 DFK 5  & 18:39:08.10  & -06:05:24.4  & 0.192 \\
Stephenson 2 DFK 6  & 18:39:18.40  & -06:00:38.4  & 0.320 \\
Stephenson 2 DFK 8  & 18:39:19.90  & -06:01:48.1  & 0.325 \\
Stephenson 2 DFK 9  & 18:39:06.80  & -06:03:20.3  & 0.332 \\
Stephenson 2 DFK 10 & 18:39:14.70  & -06:01:36.6  & 0.338 \\
Stephenson 2 DFK 11 & 18:39:18.30  & -06:02:14.3  & 0.331 \\
Stephenson 2 DFK 13 & 18:39:17.70  & -06:04:02.5  & 0.337 \\
Stephenson 2 DFK 14 & 18:39:20.40  & -06:01:42.6  & 0.339 \\
Stephenson 2 DFK 15 & 18:39:22.40  & -06:01:50.1  & 0.351 \\
Stephenson 2 DFK 16 & 18:39:24.00  & -06:03:07.3  & 0.352 \\
Stephenson 2 DFK 17 & 18:39:15.10  & -06:05:19.1  & 0.342 \\
Stephenson 2 DFK 18 & 18:39:22.50  & -06:00:08.4  & 0.363 \\
Stephenson 2 DFK 19 & 18:39:19.50  & -05:59:19.4  & 0.371 \\
Stephenson 2 DFK 20 & 18:39:24.10  & -06:00:22.8  & 0.378 \\
Stephenson 2 DFK 21 & 18:39:15.80  & -06:02:05.5  & 0.369 \\
Stephenson 2 DFK 23 & 18:39:01.50  & -06:00:59.9  & 0.396 \\
Stephenson 2 DFK 26 & 18:39:35.10  & -05:59:15.8  & 0.410 \\
Stephenson 2 DFK 27 & 18:39:16.00  & -06:05:03.2  & 0.431 \\
Stephenson 2 DFK 29 & 18:39:22.20  & -06:02:14.7  & 0.460 \\
Stephenson 2 DFK 30 & 18:39:23.40  & -05:59:01.3  & 0.489 \\
Stephenson 2 DFK 31 & 18:39:09.30  & -06:01:06.9  & 0.527 \\
Stephenson 2 DFK 52 & 18:39:23.40  & -06:02:15.9  & 0.571 \\
Stephenson 2 DFK 72 & 18:39:16.20  & -06:03:07.2  & 0.624 \\
\multicolumn{4}{c}{\citet[RSG3 cluster]{clark09}}\\
CND2009 S1  & 18:45:23.60  & -03:24:13.9  & 0.316 \\ 
CND2009 S2  & 18:45:26.54  & -03:23:35.4  & 0.196 \\ 
CND2009 S3  & 18:45:24.35  & -03:22:42.1  & 0.176 \\ 
CND2009 S4  & 18:45:25.32  & -03:23:01.1  & 0.155 \\ 
CND2009 S5  & 18:45:23.27  & -03:23:44.1  & 0.313 \\ 
CND2009 S7  & 18:45:24.18  & -03:23:47.4  & 0.238 \\ 
CND2009 S8  & 18:45:20.06  & -03:22:47.2  & 0.309 \\ 
CND2009 S9  & 18:45:28.13  & -03:22:54.6  & 0.303 \\ 
CND2009 S14 & 18:45:29.03  & -03:22:37.4  & 0.306 \\ 
2MASS J18452254-0322261 & 18:45:22.54  & -03:22:26.1  & 0.307 \\ 
\multicolumn{4}{c}{\citet[NGC 7419]{caron03}}\\
NGC 7419 BMD 435     & 22:54:16.14  & +60:49:29.0  & 0.138 \\ 
NGC 7419 BMD 696     & 22:54:22.67  & +60:48:31.7  & 0.146 \\ 
NGC 7419 BMD 921     & 22:54:30.49  & +60:47:50.7  & 0.149 \\ 
NGC 7419 BMD 139     & 22:54:01.28  & +60:47:42.0  & 0.154 \\ 
\multicolumn{4}{c}{\citet[Per OB1]{pierce00,mermilliod08}}\\
FZ Per (NGC 884 MMU 1818)   & 02:20:20.60  & +57:09:31.0  & 0.158 \\
RS Per (NGC 884 2417)    & 02:22:24.30  & +57:06:34.0  & 0.153 \\
AD Per (NGC 884 MMU 1655)   & 02:20:29.00  & +56:59:35.0  & 0.151 \\
GP Cas     	  & 02:39:50.40  & +59:35:51.0  & 0.156 \\
T Per             & 02:19:21.90  & +58:57:40.0  & 0.152 \\
YZ Per            & 02:38:25.40  & +57:02:46.0  & 0.150 \\
BU Per (NGC 869 MMU 899)    & 02:18:53.30  & +57:25:17.0  & 0.140 \\
V500Cas           & 02:51:24.00  & +57:50:00.0  & 0.138 \\
XX Per            & 02:03:09.40  & +55:13:57.0  & 0.143 \\
SU Per            & 02:22:06.90  & +56:36:14.0  & 0.111 \\
V466 Cas (NGC 457 25) & 01:19:53.64  & +58:18:31.2  & 0.138 \\
\multicolumn{4}{c}{\citet[GLIMPSE9]{messineo09}}\\
MFD2010 1 & 18:34:09.27  & -09:14:00.7  & 0.159 \\   
MFD2010 5 & 18:34:09.87  & -09:14:23.3  & 0.166 \\   
MFD2010 6 & 18:34:10.37  & -09:13:49.5  & 0.180 \\   
MFD2010 8 & 18:34:10.36  & -09:13:52.9  & 0.160 \\   
\multicolumn{4}{c}{\citet[SGR 1900+14]{davies10}}\\
VLH96 A \& VLH96 B  & 19:07:15.35  & +09:19:21.4  & 0.165 \\
\multicolumn{4}{c}{\citet[MFD2008]{messineo08}}\\
MFD2008 1    & 18:13:22.26  & -17:54:15.6  & 0.165 \\
\multicolumn{4}{c}{\citet[Galactic center]{blum03}}\\
BSD96 1     & 17:45:36.93  & -29:00:30.2  & 0.210 \\
BSD96 14    & 17:45:38.52  & -28:59:56.8  & 0.206 \\
BSD96 48    & 17:45:39.69  & -29:00:54.2  & 0.210 \\
BSD96 66    & 17:45:39.99  & -29:00:22.2  & 0.217 \\
BSD96 79    & 17:45:40.14  & -28:59:39.5  & 0.198 \\
BSD96 108   & 17:45:41.08  & -29:00:47.9  & 0.205 \\
BSD96 109   & 17:45:41.12  & -29:00:39.6  & 0.216 \\
\multicolumn{4}{c}{\citet[RSGs in the inner Galaxy]{comeron04}}\\
CTC2004 64      & 18:07:50.72  & -20:04:09.2  & 0.155 \\
CTC2004 70      & 18:08:37.35  & -19:50:05.3  & 0.152 \\
CTC2004 74      & 18:09:37.50  & -19:28:27.1  & 0.151 \\
CTC2004 97      & 18:14:53.10  & -17:00:51.3  & 0.147 \\
CTC2004 121     & 18:16:38.18  & -16:23:34.0  & 0.151 \\
CTC2004 125     & 18:16:55.97  & -16:09:54.3  & 0.157 \\
CTC2004 152     & 18:18:05.31  & -15:57:19.4  & 0.160 \\
CTC2004 184     & 18:25:11.20  & -12:28:40.2  & 0.154 \\
CTC2004 194     & 18:27:39.07  & -11:39:23.0  & 0.161 \\
\hline
\end{longtable}
\small
Columns are (from left to right) designations of the targeted RSGs (names are taken from SIMBAD), right ascension, declination, and rms noise.
}

\end{document}